\documentclass{aa}  
\usepackage{graphicx}
\usepackage{natbib}
\bibpunct{(}{)}{;}{a}{}{,}
\pdfoutput=1
\usepackage{txfonts}
\begin{document}
%

  \title{Resolving the innermost parsec of Centaurus A at mid-infrared wavelengths\footnote{Based on observations made with the Very Large Telescope Interferometer at the European Southern Observatory on Cerro Paranal.}
  }

   \author{K. Meisenheimer\inst{1}
          \and K. R. W. Tristram\inst{1}
          \and W. Jaffe\inst{2}
          \and F. Israel\inst{2}
          \and N. Neumayer\inst{1}
          \and D. Raban\inst{2}
          \and H. R\"ottgering\inst{2}
          \and W. D. Cotton \inst{3}
          \and U. Graser \inst{1}
          \and Th. Henning \inst{1}
          \and Ch. Leinert \inst{1}
          \and B. Lopez \inst{4}
          \and G.~Perrin \inst{5}
          \and A. Prieto \inst{1}
                   }

   \offprints{K. Meisenheimer, meisenheimer@mpia.de}

   \institute{Max-Planck-Institut f\"ur Astronomie, K\"onigstuhl 17, 
   D-69117 Heidelberg, Germany
         \and Sterrewacht Leiden, Niels-Bohr-Weg 2, 2300 CA Leiden, The Netherlands 
         \and NRAO, 520 Edgemont Road, Charlottsville, VA 22903-2475, USA
         \and Observatoire de la C\^ote d'Azur, Boulevard de l'Observatoire, BP 4229, 06304 Nice Cedex 4, France 
         \and Observatoire de Paris, LESIA, UMR 8109, 92190 Meudon, France
             }

   \date{Received December 19, 2006; accepted May 18, 2007}

 
  \abstract
  {} 
   {To reveal the origin of mid-infrared radiation from the core
   of Centaurus A, we carried out interferometric observations with the MID-infrared Interferometer (MIDI) at ESO's VLTI telescope array.}
   {Observations were obtained with four baselines between unit telescopes of the VLTI, two of them roughly along the radio axis and two orthogonal to it. The interferometric measurements are spectrally resolved with $\lambda/\Delta \lambda = 30$ in the wavelength range 8 to 13 $\mu$m. Their resolution reaches 15\,mas at the shortest wavelengths. Supplementary observations were obtained in the near-infrared with the adaptive optics instrument NACO, and at mm wavelengths with SEST and JCMT.}
   {The mid-infrared emission from the core of Centaurus A is dominated by an unresolved point source ($<10$\,mas). Observations with baselines orientated perpendicular to the radio jet reveal an extended component which can be interpreted as a geometrically thin, dusty disk, the axis of which is aligned  with the radio jet. Its diameter is about 0.6\,pc. It contributes between 20\% (at $\lambda \simeq 8\mu$m) and 40\% (at $\lambda \simeq 13\mu$m) to the nuclear flux from Centaurus A and contains dust at about 240 K.
We argue, that the unresolved emission is dominated by a synchrotron source. Its overall spectrum is characterized by an $F_\nu \sim \nu^{-0.36}$ power-law which cuts off exponentially towards high frequencies at  $\nu_c = 8\:10^{13}$\,Hz and becomes optically thick at $\nu < \nu_1 \simeq 45$\,GHz. 
Based on a
Synchrotron Self Compton (SSC) interpretation for the $\gamma-$ray emission,  we find a magnetic field strength of 26\,$\mu$T and a maximum energy of relativistic electrons of $\gamma_c = E_c/m_ec^2 = 8500$. Near $\gamma_c$, the acceleration time scale is $\tau_{acc} = 4$ days, in good agreement with the fastest flux variations, observed at X-ray frequencies. 
Our SSC model argues for a Doppler factor $\delta \simeq 1$ which -- together with the jet-counter jet ratio of the radio jets on parsec scale -- results in an upper limit for the bulk Lorentz factor $\Gamma_{jet} < 2.5$, at variance with the concept of a "mis-directed BL Lac object".
%
We estimate a thermal luminosity of the core, $P_{th} \simeq 1.3\;10^{34}\,{\rm W} = 1.5\;10^{-4}\times L_{Edd}$, intermediate between the values for highly efficiently accreting AGN ({\it e.g.} Seyfert galaxies) and those of typical FR I radio galaxies. This luminosity, which is predominantly released in X--rays, is most likely generated in an Advection Dominated Accretion Flow (ADAF) and seems just sufficient to heat the dusty disk.   }
{}

   \keywords{galaxies (individual) --
                Active Galactic Nuclei --
               synchrotron radiation --
               techniques: interferometric
               }
\titlerunning{The innermost parsec of Centaurus A}
   \maketitle
%

Centaurus A (NGC 5128) is the closest active galaxy. Its activity was first noticed at radio frequencies \citep{Bolton_etal49} where it is one of the brightest and largest objects in the sky, extending over about $8\degr \times 3\degr$ \citep{Junkes_etal93}. An inner system of radio jets and lobes, about 12\arcmin\ in size \citep{Clarke_etal92} has also been detected in X-rays \citep{Doebereiner_etal96, Hardcastle_CenA_2003}. The source of this large scale activity is an Active Galactic Nucleus (AGN) in the center of an elliptical galaxy, which is undergoing late stages of a merger event with a spiral galaxy \citep{BaadeMinkowski1954}. The core of Centaurus A is heavily obscured by the dust lane of the spiral and becomes visible only at wavelengths longwards of $0.8\,\mu$m \citep{Schreier_etal98, Marconi_etal00}. It harbors a super-massive black hole, the mass of which has recently been determined from the
velocity field in a circum-nuclear gas disk to be $M = 6\times 10^7 M_{\sun}$
\citep{Nadine_etal2006}. See \citet{Israel_1998} for a comprehensive review of general properties of Centaurus A . 

Centaurus A's close distance of only 3.84\,Mpc \citep{Rejkuba_2004}\footnote{Recent distance measurement of Centaurus A range between 3.4 and 4.2 Mpc with typical uncertainties of $\pm 10\%$.}  offers unique opportunities to look into the very core of an AGN, as 1 parsec corresponds to 53 mas (milli-arcseconds). Despite this fact, single-telescope observations have not been able to resolve the core at any wavelength: at short wavelengths ($\sim 1\,\mu$m) the upper limit for its size is about 100 mas (1.9 pc). Radio interferometry with VLBI networks reveals a core - jet structure within the central parsec: the well-collimated radio jet can be traced over $ > 60$\,mas at $\lambda=6$\,cm \citep{Tingay_etal98}. Also a counter-jet is clearly detected.
Nevertheless, the radio core (with inverted spectrum $S_\nu \sim \nu^{2}$ between 6 and 1 cm) is hardly resolved even with millimeter VLBI: at 43 GHz \citet{Kellerman_etal97} find an angular diameter (FWHM) of $0.5\pm 0.1$ mas (0.01 pc).

\begin{table*}
\centering          
\begin{minipage}[t]{\textwidth}
\caption{Log of the MIDI observations of Centaurus A and the calibrator HD\,112213. The derived correlated flux (averaged over $\pm 0.2\,\mu$m) is given for two wavelengths, least affected by the silicate absorption.}             
\label{t:obslog}      
\renewcommand{\footnoterule}{}  
\begin{tabular}{lrrrccc}     
\hline\hline      
& & & & & & \\[-2ex]                 
Date: Telescopes & LST & \multicolumn{2}{c}{Baseline}  &
Resolution\footnote{Resolution $\lambda/2B$ at $\lambda = 8.3\,\mu$m.}
 & \multicolumn{2}{c}{Correlated flux}\\ 
Target  & &length & P.A. & & $\lambda = 8.3\,\mu$m & $\lambda = 12.6\,\mu$m \\
             & [hour]     & [m]      & [degree]       &
  [mas]    & [Jy]  & [Jy] \\
\hline    
& & & & & & \\[-2ex]                
28-Feb-2005: UT3--UT4  & & & & & & \\
\hline
Cen A      &  12:02 & 58.2 &  96 & 14.8 & $0.45\pm 0.04$ &
 $0.73\pm 0.05$ \\  
HD\,112213 & 12:26 & 61.3 &  105 & 14.0 & & \\  
Cen A      & 14.24 & 62.4 & 120 & 13.6 & 
$0.34\pm 0.05$ & $0.60\pm 0.05$ \\  
HD\,112213 & 14:50 & 61.1 &  130 & 13.9 & & \\  
& & & & & & \\[-1ex]                 
26-May-2005: UT2--UT3  & & & & & & \\
\hline
Cen A      & 11:53 & 46.5 &  27 & 18.2 & 
$0.66\pm 0.04$ & $0.91\pm 0.06$ \\  
HD\,112213 & 12:44 & 45.5 &  40 & 17.7 & & \\  
Cen A      & 14:05 & 44.1 &  46 & 19.2 & 
$0.76\pm 0.04$ & $1.01\pm 0.06$ \\  
HD\,112213 & 14:32 & 41.5 &  54 & 20.4 & & \\[0.3ex] 
\hline                  
\end{tabular}
\end{minipage}
\end{table*}

The nature of the near- and mid-infrared emission from the parsec-size core of Centaurus A remains a matter of debate. Although several authors ({\it e.g.} \citet{Bailey_etal1986,Turner_etal1992,ChCaCe_2001}) have argued that high frequency synchrotron radiation might be an important contribution to the emission, others assume that the extremely red near-infrared colors of the unresolved core hint to the existence of a hot, AGN-heated dust structure  \citep[see][ and references therein]{Israel_1998}, which has been postulated to exist in the central parsec of more luminous AGN and has recently been resolved by mid-infrared interferometry of nearby Seyfert 2 galaxies \citep{Jaffe_etal04, Ratzka_spie2006,Tristram_etal2006}. In order to resolve the core emission on sub-parsec scales, interferometric observations are mandatory. In this paper we will report on the first high-frequency interferometry of Centaurus A, which was  obtained with the Very Large Telescope Interferometer (VLTI) of the European Southern Observatory (ESO).

\section{Observations}
\label{s:obs}
 
\subsection{Interferometric observations with MIDI}
\label{ss:obsMIDI}

The interferometric observations of Centaurus A were obtained with the MID-infrared Interferometric instrument (MIDI) at the Very Large Telescope Interferometer (VLTI) during two nights of guaranteed time on Febuary 28 and May 26, 2005
(see log of the interferometric observations in Table~\ref{t:obslog}).
We used two telescope combinations with roughly orthogonal configuration:
UT3--UT4 and UT2--UT3. 


MIDI is a classical Michelson type stellar interferometer combining the beams of two $8 \: \textrm{m}$ unit telescopes (UTs) of the VLTI in the N-band. By insertion of a NaCl prism into the light path, the instrument produces spectrally dispersed fringes from 7.8 to $13.2 \: \mu \textrm{m}$ with a spectral resolution of $R \sim 30$ \citep{Leinert_apss2003,Morel_spie2004}. At both telescopes the wavefront was corrected using the adaptive optics system MACAO \citep{Arsenault_spie2003}.
We adopted the following observing procedure:

First, MIDI was used in imaging mode to center the object on the detector, thus ensuring an overlap of the two incoming beams for the interferometric measurement. Chopping of the UT secondaries removes the sky background (chopping frequency $f = 2 \: \textrm{Hz}$, position angle $\alpha = 0\degr$ and chopping throw $\delta = 15\arcsec$).  Our experience is, that for weak targets this imaging is the most challenging part of the observation as background gradients hamper the detection of faint sources.\footnote{This will improve with the introduction of Variable Curvature Mirrors (VCMs) which were not available during our observations.} The short wave N band filter at $8.7 \: \mu \textrm{m}$ and an exposure time of $4 \: \textrm{ms}$ were used for the acquisition. To obtain a clear detection of the nucleus of Centaurus A a total of 3000 to 5000 exposures had to be taken.  

For the interferometric observations the beam combiner, a $0.6\arcsec \times 2\arcsec$ slit and the NaCl prism were inserted into the light path, resulting in two spectrally dispersed interferometric signals of opposite phase on the detector. Fringes were searched by scanning with the VLTI delay lines a few millimetres around the expected position of zero optical path difference (OPD) while MIDI's internal piezo-driven mirrors vary the OPD rapidly. No chopping is needed during interferometric measurements as the uncorrelated background signal can be removed with a software high-pass filter from the modulated fringe signal. After the fringe search had determined zero OPD, the integration in fringe tracking mode was started. In this mode the MIDI piezos change the OPD over $80 \: \mu \textrm{m}$ in order to estimate the position of zero OPD in real time from the fringe movement in every scan. 
For most fringe tracking observations the integration time per frame was ${\tt DIT} = 12$\,msec, which was increased to ${\tt DIT} = 18$\,msec for the second observation on May 26.  We took ${\tt NDIT} = 8000$ frames per fringe tracking on February 28, and ${\tt NDIT} = 5000$ on May 26. We used the offset tracking mode, 
at an offset of $50 \: \mu \textrm{m}$ from zero OPD. 

The interferometric integration was followed by two sequences of photometric data: With one shutter open, only the light from telescope A falls on the beam splitter producing two photometry signals on the detector. The integration time during photometry of Centaurus A was $12\:\textrm{ms}$, and the total number of photometry frames was increased from 4000 in the first measurement in February to 10000 frames for the measurements in May. Again chopping had to be used for the photometric measurements. The same procedure was repeated with only the shutter of telescope B open. 

After observing Centaurus A,
the entire procedure:  centering, fringe search, fringe track and photometry was repeated for the calibrator star HD\,112213, to enable a correction for atmospheric transparency and instrumental visibility in the data reduction.

\subsection{Data reduction of interferometric observations}
\label{ss:reduction}

All interferometric and photometric data were reduced with the EWS package \citep[version 1.3, see ][]{Jaffe_spie2004}. For each set of measurements essentially two spectra, the raw {\it correlated flux} $C_{corr}(\lambda)$ and the raw {\it total flux} $C_{tot}(\lambda)$ (both measured in ADU counts/s), as well as the raw {\it visibility} $V^{raw}(\lambda) \equiv C_{corr}(\lambda) / C_{tot}(\lambda)$ are determined and subsequently calibrated by using the measurements of the standard star.

To get $C_{corr}(\lambda)$ the dispersed fringe signal is extracted using a spatial weighting function ("mask") which optimizes the signal-to-noise ratio. In order to sample the point spread function of our observations ($0\farcs 6$ FWHM), we use a weighting function with an effective width of $0\farcs 70$. The two opposite phased signals are subtracted, thus removing residual background and doubling the signal amplitude. The individual data frames are phased to the same zero optical path difference (OPD) and averaged. Frames with largely discrepant OPD values are rejected.

The raw total flux $C_{tot}(\lambda)$ is extracted from the photometric frames after subtracting the (chopped) background measurements. The same spatial mask as for the interferometric measurements is used. In order to correct for any residual background, the sky value is interpolated between two sky windows running above and below the object spectrum, respectively. For the observations of Centaurus A, the best sky subtraction\footnote{That is, the most consistent $C_{tot}(\lambda)$  for all four independent measurements.} 
was obtained when using two sky windows located at 0\farcs 39 to 0\farcs 90 above and below the object spectrum, respectively. 
%
%
The raw total flux $C_{tot}$ is calculated as $\sqrt{A_{1} \cdot B_{1}} + \sqrt{A_{2} \cdot B_{2}}$ with $A_{1}$ the photometry of beam A (from the first telescope) in channel 1, $B_{1}$ the photometry of beam B (from the second telescope) in channel 1, as well as $A_{2}$ and $B_{2}$ the corresponding beams in channel 2. So defined, $C_{tot}$ equals the value of $C_{corr}$ that would be expected from the same telescopes/instrument system for a {\it point source}.  

The {\it raw visibility}, calculated as $V^{raw}(\lambda)\equiv C_{corr}(\lambda)/C_{tot)(\lambda)}$ is relatively insensitive to differences in atmospheric seeing between target and calibrator observations and is commonly used as the principal output of an interferometer.  
However it is very sensitive to photometric errors caused by background fluctuations, which are important in the mid infrared, and in some cases the direct interpretation of $C_{corr}$ is preferable.  These issues will be discussed in detail in Section\,\ref{s:results}.

  To calculate the total flux $F_{tot}$ displayed in Fig.\,\ref{f:Ftot}, we use a slightly different raw total flux $C'_{tot}(\lambda)$
which is determined as $C_{tot}(\lambda)$ but by averaging the four measurements linearly and {\it without} applying the mask: $C'_{tot} = {1 \over 4} (A_{1} + B_{1} + A_{2} + B_{2})$.
While not appropriate for calculating visibilities, this definition
of $C'_{tot}$ is less sensitive to changes in telescope pointing and
atmospheric seeing than $C_{tot}$ and thus more useful for estimating variations
in the total flux of Centaurus~A.

For the standard star of known flux and visibility, the same reduction steps lead to
raw fluxes $C^*_{corr}(\lambda)$, $C^*_{tot}(\lambda)$, $C'^*_{tot}(\lambda)$, and the raw visibility $V^{*,raw}(\lambda)$. The {\it calibrated} flux densities\footnote{In the following, we will abbreviate these flux densities as total and correlated {\it flux}, respectively.} (in Jy) of Centaurus A are then derived from the known flux $F^*(\lambda)$ of HD 112213 (spectrum based on template fit to five band photometry, van Boekel, {\it priv. comm.}) according to:\\
{\it correlated} flux density 
$F_{corr}(\lambda) = C_{corr}(\lambda) \cdot [ F^*(\lambda) / C^*_{corr}(\lambda)],$ \\
{\it total} flux density \hspace{8mm}
$F_{tot}(\lambda) = C'_{tot}(\lambda) \cdot [ F^*(\lambda) / C'^*_{tot}(\lambda)],$ \\
The calibrated {\it visibility} as a function of wavelength, is derived 
by: 
$V (\lambda) = 1 \cdot  [ V^{raw}(\lambda)/ V^{raw,*}(\lambda)], $
where HD 112213 (diameter: 2.95\,mas) is assumed to be point-like ($V^*(\lambda)\equiv 1$).

\subsection{Additional single-telescope observations}
\label{ss:obsAdd}

\subsubsection{Near-infrared photometry at $1.2 < \lambda < 2.2\,\mu$m with NACO:}

Near-infrared observations were performed on June 12 and 14, 2003, and on
April 1, 2004 with
Naos-Conica (NACO) at UT4. NACO consists of the high-resolution near-infrared imager and
spectrograph Conica \citep{Lenzen_spie1998} and the Nasmyth Adaptive Optics System (Naos) \citep{Rousset_spie1998}. It provides
adaptive-optics corrected observations in the range of 1-5 $\mu$m with
14$\arcsec$ to 54$\arcsec$ fields of view and 13 to 54 mas pixel scales.\\
The data were taken in visitor mode and seeing during observations was in the
range 0$\farcs$3-0$\farcs$8 (as measured by the seeing monitor in V-band), with
clear/photometric conditions.

There are no potential reference stars bright enough ($m_K\le$14 mag) for the
wavefront correction at a distance of $\le 30\arcsec$ to the
nucleus, necessary for a good quality of correction at
the nucleus. Therefore, we directly guided on the nucleus itself using the
unique IR wavefront sensor (WFS) implemented in Naos. This strategy provides us the
best possible wavefront correction in the vicinity of the active galactic
nucleus (AGN).
During the observations the atmospheric conditions were stable and the
performance of the IR WFS was continually very good. For 
observations in J-band we used the K-dichroic, i.e. all the nuclear
K-band light was used for the wavefront correction. While observing in K-band itself the
only possibility to achieve a good performance of the WFS was to send 90\% of
the light to NAOS and only 10\% to Conica (i.e. to use the N90C10 dichroic).
In H-band we also used the N90C10 dichroic, to get the best possible correction.

To remove bad pixels and cosmics we jittered the field on several positions on
the detector. The on-chip exposure time was 60~s in J-, 20~s in H-, and 120~s
K-band and the total exposure time 20~min in J-band, 13~min in H-, and 40~min in K-band.\\
For the flux calibration  a
separate PSF star was observed directly before and after the nucleus of Centaurus~A with the
same WFS setup and exposure time. This star was chosen from the 2MASS
point source catalogue \citep{2MASScat_2003} to match Centaurus~A's nucleus as closely as possible in angular proximity, magnitude and color.

The nucleus is unresolved at all wavelengths with a size (FWHM) of $0\farcs10$ in J-, $0\farcs088$ in H-, and $0\farcs059$ in K-band.
The flux values given in Table~\ref{t:spectrum} are extracted in circular apertures of $0\farcs 20$, $0\farcs 18$, and $0\farcs 12$ diameter in J, H and K, respectively (for details refer to Neumayer et al., in preparation). 

\subsubsection{Millimeter observations with the SEST and the JCMT:}

We determined flux densities of the Centaurus A nucleus in the
millimeter wavelength range with the 15 m Swedish-ESO Submillimetre
Telescope (SEST) on Cerro La Silla (Chile).  The measurements
presented here were obtained in February and March 2003 as close in
time as possible to the epoch of the MIDI and NACO observations
discussed in this paper. 

The SEST beamsize ranges from 57\arcsec at 85 GHz to 14\arcsec at 345 GHz.
Using scans we have determined that, at least up to 230 GHz, the continuum source is unresolved by these beams. The SEST measurements were made with a
chopping secondary in double-beamswitching mode, with a throw of
11$'$.  Because the primary aim of the observations was a study of the
absorption-line spectrum of Centaurus A, a special effort was made to get a
well-defined continuum level by frequent pointing and calibration.
Moreover, seen from Chile, the galaxy culminates at very small zenith
angles. For these reasons, the SEST measurements of the unresolved
continuum nucleus are quite accurate, as is also indicated by the
small dispersion (less than 5\% ) of individual measurements in the 3 mm
and 2 mm windows. At higher frequencies (and shorter wavelengths of
1.3 mm and 0.9 mm), both the smaller beam (making pointing more
critical) and a poorer sky transmission cause the accuracy to be
somewhat worse (typically about 15-20\% ).  The continuum levels were
measured in each individual spectrum in the velocity intervals 0-300
km/sec and 800-1100 km/sec (local standard of rest), well clear of molecular line emission centering on a systemic velocity of about 550 km/sec. We have
used these data to construct a best fit millimeter spectrum (frequency
range 85 -- 270 GHz) with spectral index ($S_\nu \sim \nu^{\alpha}$)
$\alpha = -0.41 \pm 0.05$, and extracted for each receiver band the
standardized flux densities at 90, 150 and 230 GHz listed in Table~\ref{t:spectrum}.

In March, May and July 2003, we also obtained measurements at 265/268
GHz with the 15\,m James Clerk Maxwell Telescope (JCMT) on Mauna Kea
(Hawaii), the mean of which is also listed in Table 2. The JCMT beam
at these frequencies was about 18\arcsec.  From Hawaii, Centaurus A never rises
very high in the sky, and is in fact observable only during a few
hours per day. Moreover, the observations were made in
single-beamswitch only, with a throw of 3\arcmin. The JCMT observations are
therefore less accurate (dispersion between individual scans about
20\%). In addition, we measured the nuclear flux-density in the 0.85 mm
window (330/345 GHz) a number of times in the same period  during which the MIDI and NACO observations were made (2003.30 --
2005.25). Over
the full two-year period, these JCMT measurements suggest a
significant drop in nuclear intensity from about 7 Jy to 4.5 Jy (for more details see Israel et al., in preparation).

\begin{figure}
   \centering
   \includegraphics[width=85mm]{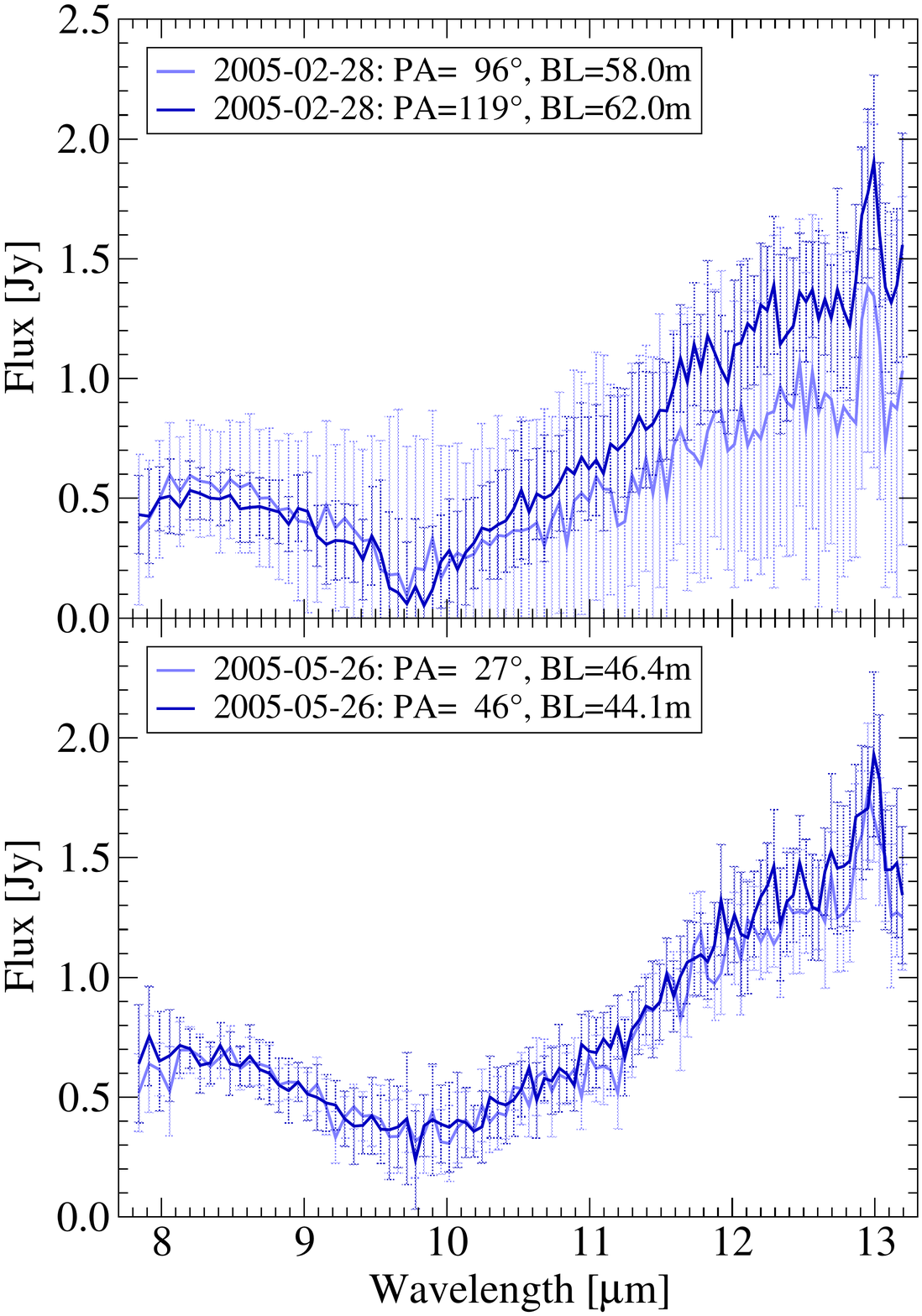}
     \caption{Spectrum of the total flux $F_{tot}$ between 8 and 13 $\mu$m as observed on Febuary 28 (top panel) and on May 26, 2005 (bottom panel). The obvious differences in $F_{tot}$ are caused by imperfect background subtraction (see text). The errors are dominated by systematic uncertainties. Note the broad silicate absorption feature at $8.5 < \lambda < 12\,\mu$m and the [NeII] emission line at $\lambda = 12.90\,\mu$m.}
\label{f:Ftot}
\end{figure}

\section{MIDI Results}
\label{s:results}

\begin{figure}
\centering
   \includegraphics[width=85mm]{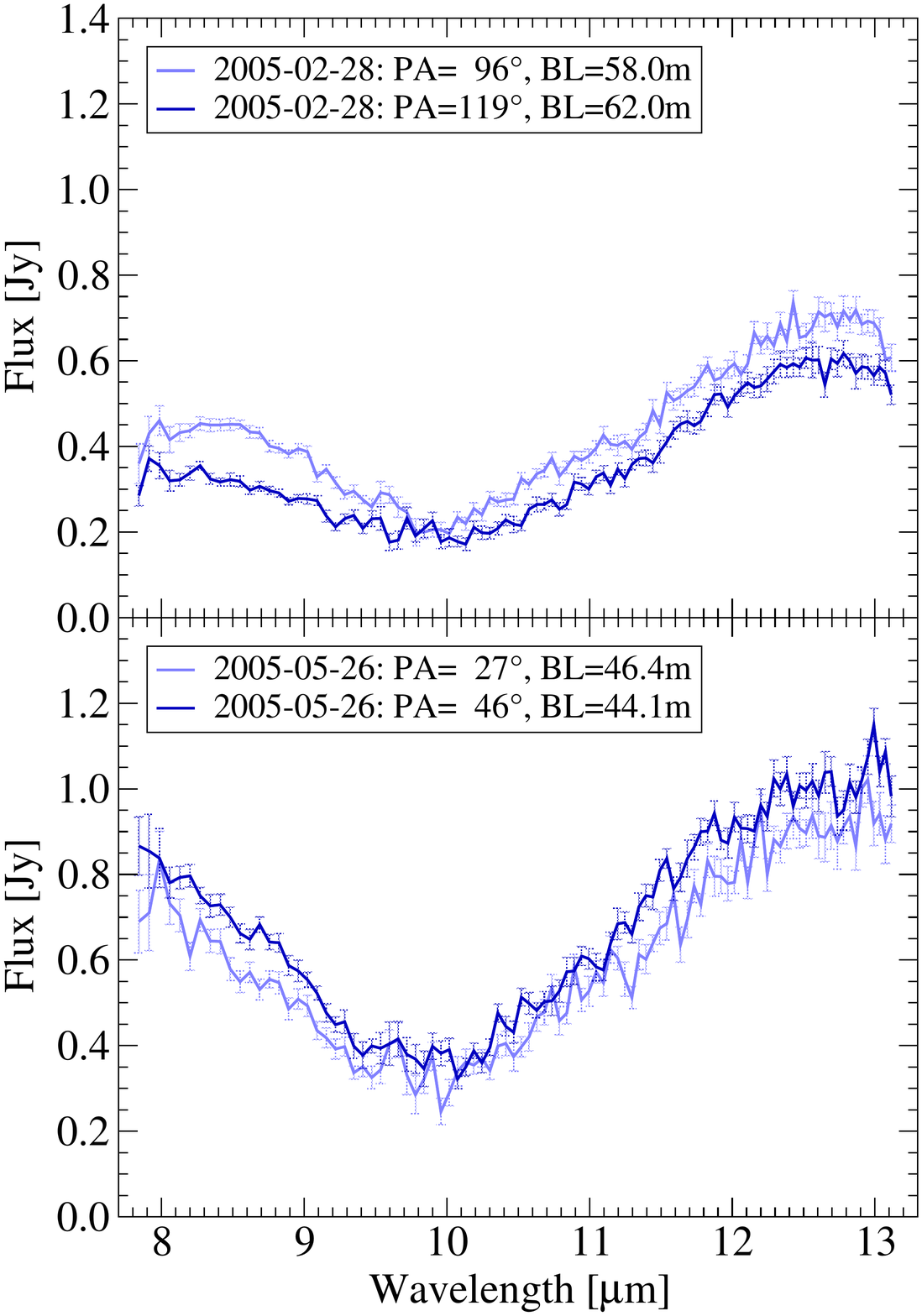}
      \caption{Spectrum of the correlated flux $F_{corr}$ between 8 and 13 $\mu$m as observed on Febuary 28 (top panel) and on May 26, 2005 (bottom panel). Wavelengths $9.5 < \lambda < 10.1\,\mu$m are affected by the atmospheric ozone band. As for $F_{tot}$ (Fig.\,\ref{f:Ftot}) the spectral shape is dominated by silicate absorption, but no evidence for the [NeII] emission line is present.
In contrast to Fig.\,\ref{f:Ftot}, here the errors are dominated by photon noise. 
}
         \label{f:Fcorr}
\end{figure}

\begin{figure}
\centering
   \includegraphics[width=85mm]{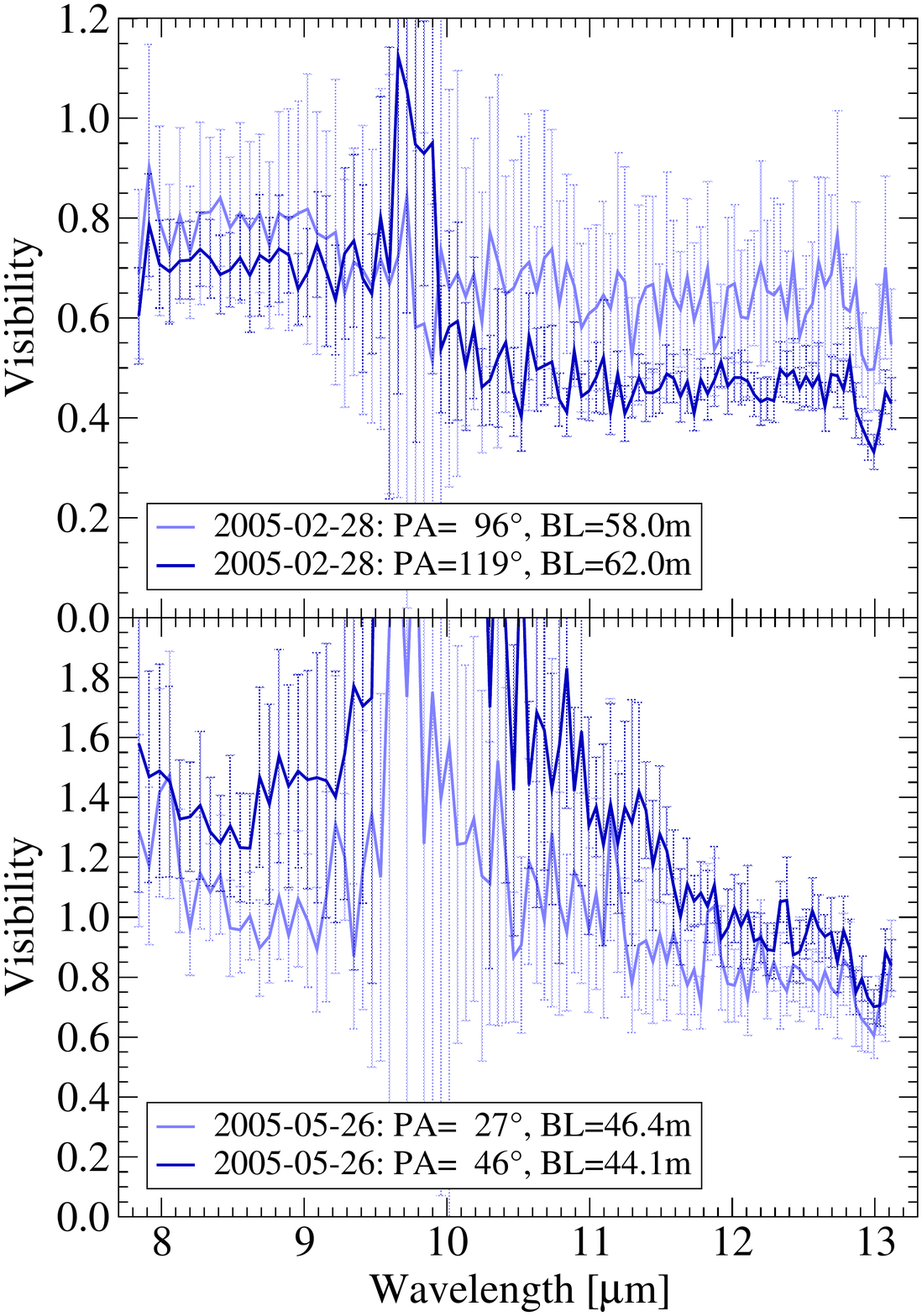}
      \caption{Spectrum of the visibility $F_{corr}/F_{tot}$ between 8 and 13 $\mu$m as observed on Febuary 28 (top panel) and on May 26, 2005 (bottom panel). Wavelengths $9.5 < \lambda < 10.1\,\mu$m are strongly affected by the atmospheric ozone band. Errors are dominated by the (systematic) errors in $F_{tot}$.}
         \label{f:Vis}
\end{figure}

The results of our MIDI observations are summarized in Figures \ref{f:Ftot} to \ref{f:Vis} which show the {\it total} flux $F_{tot}(\lambda)$, the {\it correlated} flux $F_{corr}(\lambda)$, and the {\it visibility} $V(\lambda)$ between 8 and 13 $\mu$m as observed on Febuary 28 and on May 26, 2005. 
Most of the observed spectral region is affected by the very broad absorption band due to silicates. The depth of the silicate feature is identical in $F_{corr}$ and $F_{tot}$, indicating that both the core and extended
components suffer the same extinction. The [NeII] emission line at $\lambda = 12.90\,\mu$m is clearly detected in all four spectra displayed in Fig.\,\ref{f:Ftot}, but not present in any of the correlated flux spectra in Fig.\,\ref{f:Fcorr} which have superior signal-to-noise ratio. This indicates that the [NeII] emission line arises in an extended region ($> 50$\,mas), which is over-resolved by the interferometric observations.

Centaurus A was one of the first targets to be observed with MIDI with an average N-band flux $\langle F_{tot} \rangle$ below 1 Jy. For such faint sources it is a greater challenge to measure the total flux $F_{tot}$ accurately, rather than to determine the correlated flux, as the strong background naturally cancels out in the interferometric observations. Although Fig.\,\ref{f:Ftot} shows that we managed to get largely consistent results for $F_{tot}$ in the two epochs, a closer inspection reveals discrepancies of 30\% around 12$\,\mu$m during one night (top panel: February 28). When comparing both nights we find deviations $> 35\%$ in the silicate absorption feature (compare
top and bottom panel) which can increase to more than a factor of 3 in the atmospheric ozone absorption band between 9.5 and 10.0$\,\mu$m . We attribute these discrepancies to uncertainties in the
background subtraction. 

An estimate of the uncertainties in determining the
correlated flux $F_{corr}$ (see Fig.\,\ref{f:Fcorr}) might be obtained by comparing the measurements over one night ({\it i.e.} with similar baseline, see Table~\ref{t:obslog}). Over most of the spectral range they are confined to
$\pm 10\%$. However, the significant difference between the measurements of $F_{corr}$ on February 28 and on May 26 have to be interpreted as true interferometric signal, showing that the core of Centaurus A is marginally
resolved along $P.A. \simeq 120\degr$ with a 60\,m baseline. 

Due to the afore mentioned uncertainties in $F_{tot}$ it is hard to judge which of the details
observed in the spectral visibilities $V(\lambda) = F_{corr} /F_{tot}$ displayed in Fig.\,\ref{f:Vis} are real: clearly the values $V(\lambda) > 1$
obtained from the May 26 observations (lower panel) are caused by incorrect background subtraction in $F_{tot}$ and thus indicate that the level of uncertainty in the visiblity measurements can reach 30\%. 
Accordingly, we regard
the two visibility measurements of February 28 (top panel) as consistent with each other despite the change in baseline position angle by 23\degr . 

\begin{figure}
\centering
   \includegraphics[width=85mm]{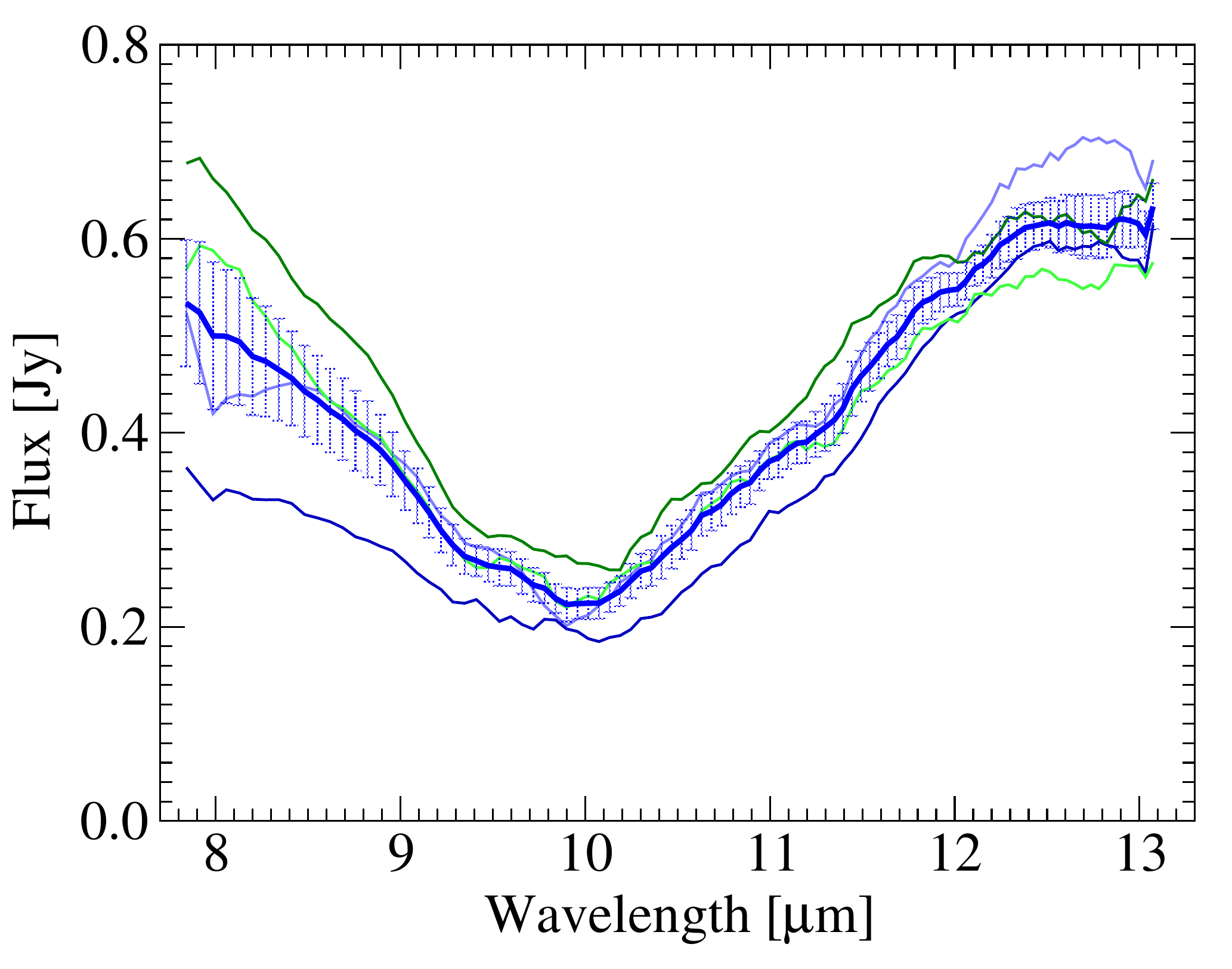}
     \caption{Spectrum of the four individual measurements and the averaged correlated flux $F_{corr}$ of the unresolved core (see text for details). The errors are derived from the scatter between the individual spectra. }        \label{f:MIDIcore}
\end{figure}

Taking all discussed uncertainties into account we conclude that there is no indication for intrinsic flux variability between the two observed epochs and
that the most robust result of our measurements is the difference in correlated flux between the two observations (that is projected baselines). The ratio 
$$f_{12} \equiv { F_{corr} (Feb28) \over F_{corr} (May26) }$$
can be approximated by a linear function 
$f_{12}(\lambda) = 0.8 - 0.04\, (\lambda - 8\,\mu{\rm m})$ between 8 and 13$\,\mu$m (compare Fig.\,\ref{f:MIDIcore}). 
As on May 26 we find $F_{corr} \equiv F_{tot}$ within the errors, we regard $V^{May26}(\lambda) \simeq 1$ and $V^{Feb28}(\lambda) = f_{12}(\lambda)$ as best measurements of the visibilities. The decrease in visibility towards longer wavelengths indicates that the source is significantly extended along $P.A. \simeq 120\degr$ and the extended emission has a spectrum which rises steeply between 9 and 13$\,\mu$m, as expected for emission from thermal dust at temperatures $T < 300$\,K (see detailed discussion in \,\ref{ss:thermaldust}). We derive the spectrum of the compact core (Fig.\,\ref{f:MIDIcore}) by averaging $F_{corr}(Feb28)$ and $f_{12} F_{corr}(May26)$. From the formal 2$\sigma$-limit of the visibility around $\lambda = 8\,\mu$m observed on May 26 ($V \ge 0.9 $), one derives an upper limit of about 6\,mas FWHM for the size of the core.

\section{The overall core spectrum of Centaurus A}
\label{s:spectrum}

In order to understand the nature of the -- unresolved -- core emission between 8 and 13$\,\mu$m ($\nu = 2.3 \dots 3.7\:10^{13}$\,Hz) it is necessary to consider not only our interferometric measurements with MIDI but also our photometry at lower and higher frequencies, as well as supplementary data from the literature. At radio frequencies ($\nu < 43$\,GHz, $\lambda > 7$\,mm),
this is straightforward as the VLBA clearly outperforms our mid-infrared
interferometry in terms of resolution, and extinction is not an issue.
Table\,\ref{t:spectrum} lists interferometric measurements of the core flux based on the VLBI and VLBA maps by \citet{Tingay_etal98}. The spectrum of the radio core is strongly inverted, $\alpha \ga 2$ for $F_\nu \sim \nu^\alpha$ (compare Fig.\,\ref{f:spectrum}). It is unresolved at $\nu \le 22$\,GHz but has been marginally resolved at 43\,GHz \citep[$0.5\pm 0.1$ mas FWHM;][]{Kellerman_etal97}. 

\begin{table*}
\begin{minipage}[t]{\textwidth}
\caption{Flux measurement of the core of Centaurus A.}             
\label{t:spectrum}      
\centering          
\renewcommand{\footnoterule}{}  
\begin{tabular}{crclcccl}     
\hline\hline      
& & & & & & & \\[-2ex]                 
Frequency & Wavelength & $F_\nu$ & $F_{\nu,0}$\footnote{Corrected for extinction by adopting $A_V = 14$\,mag (see text).} &
Date\footnote{In cases in which several observations were averaged, we
give an average date.}   & 
Instrument & Beamwidth & Reference \\
${\rm [Hz]}$  & [$\mu$m]       & [Jy]     & [Jy]  &  [year] &  & [mas]    & \\
\hline    
& & & & & & & \\[-2ex]                
$4.8\:10^9$  & 63 000 & 1.2 &   &    
1993.13 & VLBI & 2.6 & Tingay et al. 1998  \\
$8.4\:10^9$  & 35 700 & 2.4 &   &    
1996.22 & VLBA & 2.4 & Tingay et al. 1998  \\
$22.2\:10^{9}$ & 13 500 & 3.5 &   &    
1995.88 & VLBA & 1.2 & Tingay et al. 1998  \\
$90.0\:10^{9}$ & 3 530  & $8.6\pm0.6$ &   &    
2003.18 & SEST & 57 000 & this paper \\
$1.50\:10^{11}$ & 2 000 & $6.9\pm0.3$ &   &    
2003.18 & SEST & 32 000 & this paper  \\
$2.35\:10^{11}$ & 1 270 & $5.8\pm0.2$ &   &   
2003.18 & SEST & 20 000 & this paper  \\
$2.70\:10^{11}$ & 1 110 & $5.9\pm1.0$ &   &   
2003.30 & JCMT & 18 000 & this paper  \\
$3.75\:10^{11}$ & 800   & 8.5 &   &    
1991.35 & JCMT & 14 000 & Hawarden et. al. 1993 \\
$6.67\:10^{11}$ & 450   & 6.3 &   &    
1991.35 & JCMT & 10 000 & Hawarden et. al. 1993 \\
$2.38\:10^{13}$ & 12.6 & $0.62\pm0.03$ & 1.125  &    
2005.28 & MIDI & 22 & this paper  \\
$2.63\:10^{13}$ & 11.4 & $0.43\pm0.03$ & 1.074  &    
2005.28 & MIDI & 20 & this paper  \\
$2.88\:10^{13}$ & 10.4 & $0.25\pm0.02$ & 0.987  &    
2005.28 & MIDI & 17 & this paper  \\
$3.23\:10^{13}$ & 9.3 & $0.28\pm0.02$ & 1.135  &    
2005.28 & MIDI & 17 & this paper  \\
$3.61\:10^{13}$ & 8.3 & $0.47\pm0.05$ &  0.869 &    
2005.28 & MIDI & 14 & this paper  \\
$7.90\:10^{13}$ & 3.80 & $0.20\pm0.04$ & 0.368  &    
2003.36 & NACO & 90 & Prieto, priv. comm. \\
$1.35\:10^{14}$ & 2.22 & $41.5\:10^{-3}$ & 0.190  &    
1997.61 & NICMOS & 250 & Marconi et al. 2000  \\
$1.39\:10^{14}$ & 2.15 & $(33.7\pm2.0)\:10^{-3}$ & 0.169  &    
2004.25 & NACO & 59 & this paper \\
$1.80\:10^{14}$ & 1.67 & $(4.5\pm0.3)\:10^{-3}$ & 0.052  &    
2003.45 & NACO & 88 & this paper \\
$1.87\:10^{14}$ & 1.60 & $4.8\:10^{-3}$ & 0.065  &    
1997.69 & NICMOS & 170 & Marconi et al. 2000  \\
$2.34\:10^{14}$ & 1.28 & $(1.3\pm0.1)\:10^{-3}$ & 0.049  &    
2003.45 & NACO & 100 & this paper \\
$3.69\:10^{14}$ & 0.81 & $7\:10^{-6}$ & 0.010  &    
1997.80 & WFPC2 & 100 & Marconi et al. 2000  \\
[0.3ex]  
\hline                  
\end{tabular}
\end{minipage}
\end{table*}

\begin{figure}
\centering
\includegraphics[width=90mm]{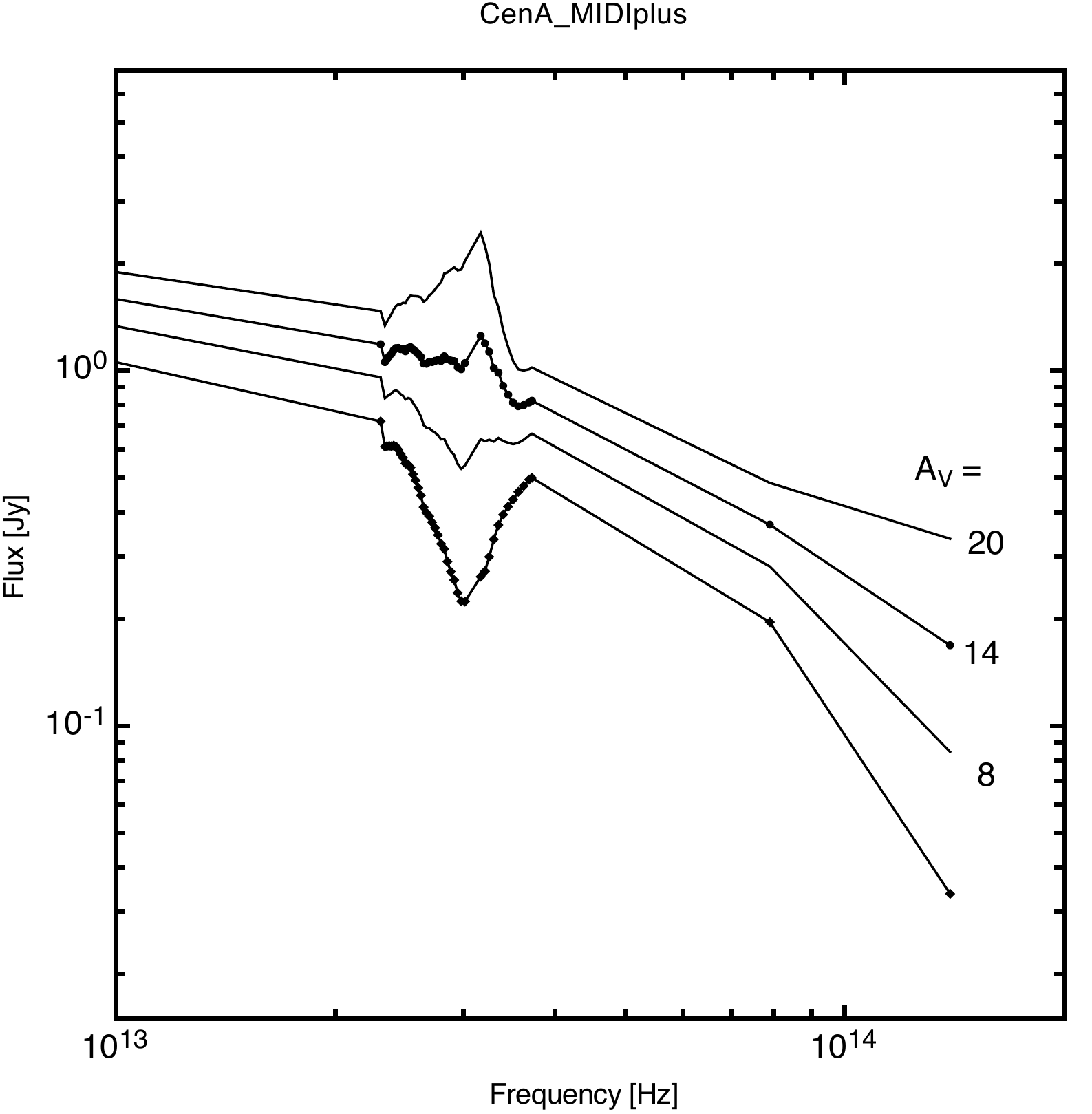} 
\caption{Spectrum of the core of Centaurus A between $10^{13}$ and $2\:10^{14}$\,Hz for different value of the assumed foreground extinction. Filled diamonds show observed flux $F_{corr}$ (averaged over all measurements), filled dots are corrected for the foreground extinction of $A_V = 14$\,mag. Remaining residuals of $\pm10\%$ at $9.5\,\mu$m$\; >\lambda > 8.2\,\mu$m are caused by an imperfect match of the short wavelength shape of the silicate absorption. Neither the assumption of minimum foreground extinction $A_V = 8$\,mag nor that of higher extinction $A_V = 20$\,mag does lead to a satisfactory removal of the silicate feature. }
\label{f:correctMIDI}
\end{figure}

To determine the core spectrum in the (sub-)mm regime between 90 and 670 GHz 
($3\,{\rm  mm} > \lambda > 0.45\,{\rm mm}$) is
much more problematic due to the lack of interferometric data and the contribution of thermal emission of cold dust in the dust lane ($T \simeq 35$\,K) shortwards of $\lambda \simeq 800\,\mu$m  \citep{Hawarden_etal93}. Nevertheless, we think that our new millimeter photometry between 90 and 270 GHz -- albeit obtained with single dish telescopes -- should represent the core flux rather well, since
the most important contaminants, the kiloparsec radio jet with its steep
spectrum $F_\nu \sim \nu^{-0.75}$ \citep{Clarke_etal92}, and the thermal dust emission, dominant at shorter wavelengths, should be negligible here.\footnote{From Fig.\,3 in \citet{Hawarden_etal93}, we estimate a maximum contamination from the dust lane of $<0.4$\,Jy ($< 10\%$) within our 18\arcsec\ beam at 270 GHz.} 
Indeed we find no deviations of our flux measurements between 90 and 270 GHz from a straight, non-thermal power-law $F_\nu \sim \nu^{-0.41}$. However, as illustrated by comparison of our photometry from 2003 with that derived 12 years earlier by \citet{Hawarden_etal93} from mapping observations at 800 and 450$\,\mu$m (see Table~\ref{t:spectrum} and Fig.\,\ref{f:spectrum}) variability is significant at these wavelengths and can reach a factor of 1.5 or more. Thus
one has to be careful when trying to reconstruct an overall spectrum from non-simultanous observations.

So far we have considered only radio to sub-mm frequencies, at which dust {\it extinction} can be neglected. This simplification certainly does not apply at $\lambda < 30\,\mu$m ($10^{13}$ Hz): there is no way to obtain the intrinsic spectrum of the core of Centaurus A without correcting for the obvious extinction on our line-of-sight. 

An absolute minimum for the extinction towards the core of Centaurus A is set by the value $A_V \simeq 8$\,mag determined from the arcsec-scale extinction map by
\citet{Marconi_etal00} and Neumayer (priv. comm.). Presumably this extinction is caused by the dust lane in Centaurus A. However, based on the presence of a circum-nuclear disk of about 80 pc diameter, observed in molecular \citep{Israel_1998} and ionized gas \citep{Schreier_etal98,Marconi_etal00}, it is expected that the total extinction on our line-of-sight towards the core is much higher. In fact, extinction values between $A_V \simeq 14$\,mag and $A_V > 40$\,mag have been discussed in the literature. Here we estimate the extinction towards the mid-infrared core by (i) assuming a galactic extinction law \citep{Schartmann_etal05} with modified silicate profile \citep[using][]{Kemper_etal2004}, and (ii) requiring the extinction corrected spectrum in the range $8\,\mu$m $< \lambda < 13\,\mu$m to be as smooth as possible, {\it i.e.} the prominent silicate feature disappears (see Fig.\,\ref{f:correctMIDI}). This leads to our "best-guess" value $A_V = (14 \pm 2)$\,mag, where the error is estimated from the fact that $A_V = 8$\,mag and $A_V = 20$\,mag are clearly rejected. Note that both our interpretation of the overall core spectrum at $\lambda < 1$\,mm as optically thin synchrotron emission (see below) and the assumption of circum-nuclear dust emission virtually exclude values $A_V > 25$\,mag since such high values would result in an erratic upturn of the intrinsic spectrum shortwards of 3$\,\mu$m. 

We list both the {\it observed} ($F_\nu$) and {\it extinction corrected} ($F_{\nu,0}$) values of the core flux in Table~\ref{t:spectrum}, and display them in Fig.\,\ref{f:spectrum} as open circles and filled dots, respectively. The five values derived from our interferometric observations represent $F_{corr}$ averaged over the measurement on February 28 and May 26, 2005.
The solid line in Fig.\,\ref{f:spectrum} gives the best-fit standard synchrotron spectrum between $4\times 10^{10}$ and $2\times 10^{14}$\,Hz. It is characterized by an optically thin power-law $F_\nu \sim \nu^{-0.36}$ which cuts off exponentially above some cutoff frequency $\nu_c = 8 \times 10^{13}$\,Hz, and becomes optically thick below $\nu_1 = (45\pm 5)$~GHz. We regard obvious discrepancies between this synchrotron spectrum, and the intrinsic, extinction corrected flux values $F_{\nu,0}$ as further evidence for variability of the core of Centaurus A. It should be noted, that for a synchrotron spectrum with high frequency cutoff one naturally expects high variability at $\nu \ga \nu_c$ since small variations in $\nu_c$ can result in large flux variations. Indeed, variations by more than a factor 3 have been observed at 3.6$\,\mu$m by \citet{Lepine_etal1984} and at 3.3$\,\mu$m by
\citet{Turner_etal1992}.

\begin{figure}
\centering
\includegraphics[width=90mm]{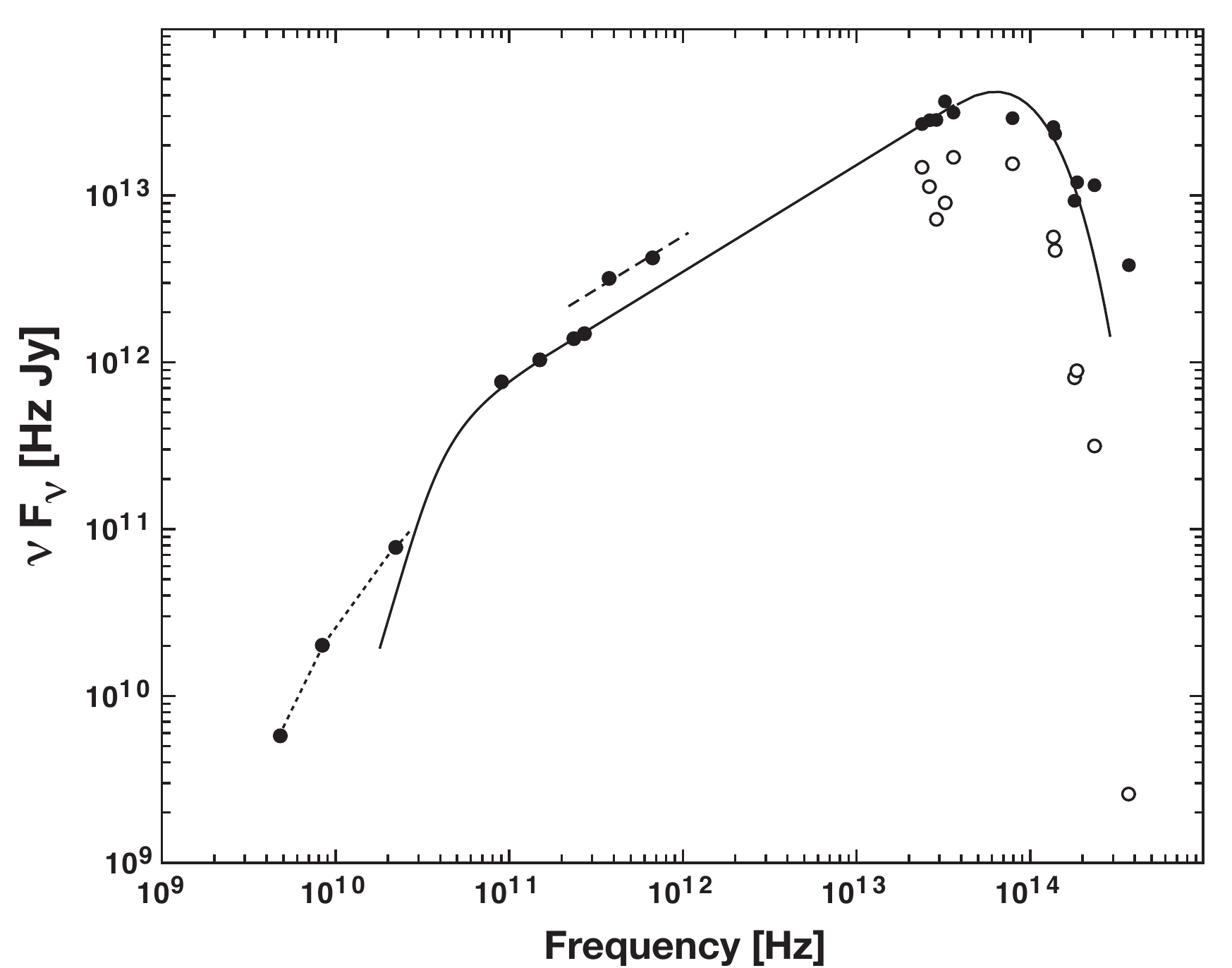} 
\caption{Overall spectrum of the core of Centaurus A. Open circles show observed flux values, filled dots are corrected for the foreground extinction of $A_V = 14$\,mag (compare Fig.\,\ref{f:correctMIDI}). The synchrotron spectrum (solid line) shows an optically thin power-law $F_\nu \sim \nu^{-0.36}$ which cuts off exponentially at $\nu_c = 8\,10^{13}$\,Hz, and is self-absorbed below $\nu_1 \simeq 4.5\,10^{10}$\,Hz. Evidence for variability exists around $3\,10^{11}$\,Hz (dashed line through photometry in 1991) and above $\nu_c$ (various epochs between 1997 and 2005, {\it cf.} Table~\ref{t:spectrum}). The excess at cm wavelengths ($\nu < 2\,10^{10}$\,Hz, connected by dotted lines) is due to optical thick components of larger size.
}
\label{f:spectrum}
\end{figure}

\section{Discussion}
\label{s:discussion}

Our interferometric MIDI observations reveal the existence of two components in the inner parsec of Centaurus A: a resolved component, the "disk", which is most extended along $P.A. \simeq 120\degr$ and the unresolved "core". In Fig.\,\ref{f:spectrum} we demonstrate that the core spectrum can be fitted by a synchrotron spectrum with millimeter-to-mid-infrared power-law $F_\nu \sim \nu^{-0.36}$ which cuts off exponentially towards higher frequencies.
However, the spatial resolution of our present observations is not sufficient
to establish unambiguously the non-thermal nature of the core emission by a surface brightness argument. Before proceeding further with this interpretation,
therefore,  it is worthwhile to consider other explanations.

\citet{Marconi_etal00} have proposed an alternative model of the near-infrared spectrum of Centaurus A, which consists of a compact, hot black body (dust at $T  = 700$\,K) plus a non-thermal power-law $ \sim \nu^{-0.9}$ which they had extrapolated from X-ray observations \citep{RXTE_1999} to lower frequencies.
In order to fit the spectrum, they had to assume that only the power-law component is reddened by $A_V \simeq 14$\,mag, while the hot dust suffers no more than the foreground extinction ($A_V = 7.8$\,mag). Furthermore, they argue that the hot blackbody could be small enough to show the observed 3.6$\,\mu$m variability.
There exist several problems with this model: first, our interferometric measurements prove that any core emission suffers at least $A_V = 14$\,mag of extinction. Second, the steep X-ray spectrum $ \sim \nu^{-0.9}$ has not been confirmed by subsequent observations with XMM and Chandra \citep{Evans_etal2004}. The extrapolation of the true X-ray spectrum leads to a
negligible contribution in the near-infrared. Also it should be noted, that exponential cut-offs are natural in synchrotron spectra and therefore no additional component is needed to explain the steep NIR spectrum of the core.  Last but not least, both the variability at $\lambda \le 3.6\,\mu$m which seems to be correlated with radio variations \citep[][ see also Fig.\,\ref{f:spectrum}]{Lepine_etal1984} and the high polarization \citep{Bailey_etal1986} are explained much more naturally in terms of a synchrotron model.

Therefore, we conclude 
that the core emission is dominated by non-thermal synchrotron radiation.
On the other hand, the "disk" emission is most naturally explained by
thermal emission of AGN heated dust at $T \simeq 300$\,K as seen in other AGN. Further MIDI observations with projected baseline $> 100$\,m (using UT1--UT4) will allow us to pin down the flux ratio between "core" and "disk" more accurately. We defer a detailed discussion of the dust emission to section\,\ref{ss:thermaldust} and start here with the discussion of the core spectrum.

The overall spectrum of the core in Centaurus A in Fig.\,\ref{f:spectrum} is reminiscent of millimeter-to-optical blazar spectra \citep[see, e.g., ][]{Bregman_1990}. This and the detection of $\gamma-$rays from Centaurus~A has led several authors to jump on the conclusion that the core spectrum provides
additional evidence for Centaurus A being a ''mis-directed BL Lac object''
\citep{Bailey_etal1986,ChCaCe_2001}. We do not want to follow this path for two reasons:
\begin{enumerate}
\item In the standard unified picture of BL Lac objects \citep{UrryPadovani_1995} normal FR I radio galaxies are the (mis-directed) parent population of highly beamed BL Lac objects. Typical FR I cores are weak and their spectra normally do not extend beyond $10^{11}$ Hz.
\item With an optically thin $F_\nu \sim \nu^{-0.36}$ spectrum in the range
between $10^{11}$ and $3\times 10^{13}$ Hz, the core
spectrum of Centaurus A is exceptionally flat. Most classical blazars
display much steeper spectra in this frequency range ($\alpha = -0.6 \dots -0.9$).\footnote{It should be noticed, however, that the exceptionally flat spectrum of Centaurus A between $10^{11}$ and $3\:10^{13}$\,Hz is not unique: the nearby BL Lac object Mkn 421 also exhibits $\alpha \simeq -0.35$ in the same frequency range \citep{Macomb_etal1995}.}
\end{enumerate}

\subsection{The synchrotron core}
\label{ss:synchrotron}
The intrinsic properties of the core synchrotron source can be derived
from observed properties and standard synchrotron theory \citep{Pacholczyk_1970}, in which the self-absorption frequency $\nu_1$ and the synchrotron luminosity $P_\nu$ can be used to disentangle the strength of the magnetic field $B$ and the number density of relativistic particles in a source of known size. In the following, we will first derive the basic properties of the synchrotron source in Centaurus~A (in \ref{ss:syn_basic}), and second discuss its relation to the radio jet (in \ref{ss:syn_derived}). 

\subsubsection{Basic properties}
\label{ss:syn_basic}

The observed properties of the synchrotron core in Centaurus A are:
\begin{enumerate}
\renewcommand{\labelenumi}{( \alph{enumi} )}
\renewcommand{\labelenumii}{( \arabic{enumii} )}
\item The size determined by the VLBA observations at 43\,GHz \citep{Kellerman_etal97}: taking their observed FWHM $= 0.5 {\rm mas} = 0.0094 {\rm pc} = 2.9\,10^{14}$\,m as diameter of a quasi-homogeneous blob, one derives a radius 
$R_{43} = 1.5\,10^{14}$\,m for the synchrotron core. Comparing this with
the Schwarzschild radius of the $M = 6\times10^7 M_{\sun}$ black hole ($R_s = 1.8\,10^{11}$\,m = 1.2 AU), one finds $R_{43} = 830\,R_s$.
\item The slope of the optically thin synchrotron emission $F_\nu \sim \nu^\alpha$ with $\alpha = -0.36\pm0.01$ corresponds to an underlying electron energy distribution
$$ n(\gamma) d\gamma \equiv n_{1000} \left({\gamma \over 1000}\right)^{-q} d\gamma $$
(where we use $\gamma \equiv E/m_ec^2$ as dimensionless electron energy) with $q = 1 - 2\alpha = 1.72$.
\end{enumerate}
Following standard synchrotron theory one can use a combination of
\begin{enumerate}
\renewcommand{\labelenumi}{( \alph{enumi} )}
\renewcommand{\labelenumii}{( \arabic{enumii} )}
\addtocounter{enumi}{2}
\item the {\it intrinsic}\footnote{Intrinsic values refer to the restframe of the synchrotron emitting source.} self-absorption frequency $\nu_1 \equiv \nu_{\tau=1} = (4.5\pm 0.5)\, 10^{10}\:\delta^{-1}$\,Hz, where 
$\delta = \sqrt{1-\beta_{jet}^2} / (1 - \beta_{jet} \cos \theta)]$ is the Doppler factor of the emitting source moving with $\beta_{jet} = v_{jet}/c$ under an angle $\theta$ with respect to the line-of-sight, and the simplification 
$$\tau = \int_0^{R}{\kappa_\nu dl} \simeq \kappa_\nu R = 1, \hspace{3mm}{\rm and}$$
\item the emitted power at some (optically thin) frequency $\nu$:
\begin{eqnarray*}
P_\nu(3\,10^{11}\,{\rm Hz})  &=& 5.4\,10^{-26}~ \delta^{\alpha-2} 4\pi D_L^2\hspace{2mm}{\rm WHz^{-1}}\\ &=& 9.53\,10^{21} \delta^{\alpha-2} \hspace{2mm}{\rm WHz^{-1}},
\end{eqnarray*}
\end{enumerate}
to solve for the average magnetic field $\langle B \rangle$ and $n_{1000}$, respectively, since:
\begin{equation}
\kappa_\nu = {1 \over R} = c_\kappa(q) \cdot n_{1000}\;
\left({\langle B \rangle \over 1{\rm mT}}\right)^{1+q/2} \left( {\nu_1 \over \nu_0} \right)^{-2-q/2}
\end{equation}
and 
\begin{equation}
\epsilon_\nu = {P_\nu \over {4\over3} \pi R^3} = c_\epsilon(q) \cdot n_{1000}\;
\left({\langle B \rangle \over 1{\rm mT}}\right)^{1/2+q/2} \left( {\nu_1 \over \nu_0} \right)^{1/2-q/2} .
\end{equation}
Here we assume a spherical source\footnote{In the absence of any structural information and in the view of the observational uncertainties in size and self-absorption frequency, this over-simplification seems appropriate.}
of radius $R$ (that is $V = {4 \over 3} \pi R^3$), homogeneously filled with a
tangled field of average (transverse) strength $\langle B \rangle$. The constants $\nu_0 = 1.254\,10^{19}$\,Hz, $c_\kappa(1.72) = 3.96\,10^{-42}$, and $c_\epsilon(1.72) = 2.08\,10^{-28}$ are taken from \citet{Pacholczyk_1970}, and
converted into our units, where necessary (note: 0.1\,mT = 1 G). If we parameterize the source radius $R$ in units of the radius $R_{43}$ derived from the VLBA measurement this yields:
\begin{equation}
 \langle B \rangle  = 46\, {\rm \mu T}\times \delta^{-1}\left({R \over R_{43}}\right)^4 ,
\end{equation}
 and
\begin{equation}
n_{1000} = 3.54\,10^5 {\rm m^{-3}}\times \delta^{-1} \left({R \over R_{43}}\right)^{4\alpha -7}
\end{equation}
The low apparent velocity and the jet/counter-jet ratio ${\cal R}_{jc} = 4\dots 8$ of the parsec-scale jet \citep{Tingay_etal98} argue for Doppler factors between $\delta=1.2$ (for $\beta=0.5, \theta = 55\degr$) and $\delta=0.6$ ($\beta=0.9, \theta = 70\degr$).  However, it should be noted that the estimate of $\langle B \rangle$ from the self-absorption frequency $\nu_1$ depends very strongly on both $\nu_1$ ($\propto \nu_1^{2\alpha-5}$)  and the source size ($\propto R^4$). It is, therefore, no more than an order of magnitude
estimate. Nevertheless, we use the field strength (3) to convert observed characteristic frequencies $\nu_1^{obs},~\nu_c^{obs}$ into electron energies:
$$ \gamma_c = (\nu_c \delta^{-1} / 4.2\, 10^4 {\rm Hz})^{1/2} 
(\langle B \rangle / 1 {\rm  \mu T} )^{-1/2} = 6400\:(R/R_{43})^{-2},$$
and
$$ \gamma(\nu_1) = 153\: (R/R_{43})^{-2}.$$

\begin{table*}
\caption{Intrinsic parameters of the synchrotron core in Centaurus A. Numerical values are given for Doppler factor $\delta = 1$ and a core radius $R_c = 1.26\,10^{14}$\,m.}             
\label{t:parameters}      
\centering                          
\begin{tabular}{l c l c l}        
\hline\hline                 
Parameter & &  & Value & Units \\    
\hline                        
Optically thin spectral index           & $\alpha$ & 
$F_\nu \sim \nu^\alpha$ & $-0.36\pm 0.01$ & \\
Self-absorption frequency & $\nu_1$  & $= \nu(\tau = 1)$ & $(4.5\pm0.5)\,10^{10}$ & Hz \\
Cutoff frequency         & $\nu_c$  & & $8.0\,10^{13}$ & Hz \\
Black hole mass          & $M_{bh}$ & & $6\,10^7$      & M$_{\sun}$ \\ 
Schwarzschild radius     & $R_S$    & & $1.8\,10^{11}$ & m \\ 
Observed half-size at 43 GHz & $R_{43}$ & & $(1.5\pm 0.3)\,10^{14}$ & m \\
\hline
Radius of synchrotron core & $R_c$    & & $1.26\,10^{14}$ & m \\
Doppler factor           & $\delta$ &
$\sqrt{1-\beta^2}/(1-\beta\cos \theta)$    & 1              & \\
that is for $\theta = 50\degr$:  & & & \\[-0.5ex]
\hspace{5mm} Velocity    & $\beta_{jet} = v_{jet}/c$ & & 0.91 & \\[-0.5ex]
\hspace{5mm} Doppler factor at $\theta =0$ & $\delta_0$ & & 4.6 & \\ 
that is for $\theta = 70\degr$:  & & & \\[-0.5ex]
\hspace{5mm} Velocity    & $\beta_{jet} = v_{jet}/c$ & & 0.61 & \\[-0.5ex]
\hspace{5mm} Doppler factor at $\theta =0$ & $\delta_0$ & & 2.0 & \\
\hline
Magnetic field strength  & $\langle B \rangle$ &
$\propto \delta^{-1}R_c^{4} $       & 26          & $\mu$T \\
Relativistic particle density & $n_{1000}$ &
$\propto \delta^{-1}R_c^{4\alpha-7}   $       & $3.73\,10^5 $ & m$^{-3}$ \\
Minimum particle energy  & $\gamma_{min} < \gamma(\nu_1)$ &
$\propto R_c^{-2} $                 & $< 204$        & $m_e c^2$ \\
Cutoff particle energy   & $\gamma_c$ &
$\propto R_c^{-2} $                 & 8500           & $m_e c^2$ \\
Field energy density     & $u_B$ &
$\propto \delta^{-2}R_c^{8} $          & $0.32\, 10^{-3} $ & J\,m$^{-3}$ \\
Particle energy density     & $u_{e\pm}$ &
$\propto \delta^{-2}R_c^{4\alpha-7} $          & $0.24\, 10^{-3} $ & J\,m$^{-3}$ \\
Radiation energy density     & $u_{syn}$ &
$\propto \delta^{\alpha-2}R_c^{-2} $   & $0.98\,10^{-3}$ & J\,m$^{-3}$ \\
Synchrotron luminosity       & $P_{syn}$ &
$\propto \delta^{\alpha-2}$              & $6.8\,10^{34}$ & W \\
Acceleration time scale  & $\tau_{acc}(\gamma_c)$  &  & 4.0 & days \\
\hline                                   
\end{tabular}
\end{table*}

\noindent
Additionally, the energy density of the magnetic field within the source can be estimated:
$$ u_B = 1.0\:10^{-3} ~\delta^{-2}(R/R_{43})^{8}~ {\rm J\,m^{-3}},$$
which obviously depends very steeply on the assumed source radius $R \le R_{43}$. In any case $u_B$ is much higher than the radiation energy density of the CMB or the starlight in the core of Cen~A. As the synchrotron luminosity $P_{syn} = \delta^{\alpha-2} \int_{\nu_1}^{{1 \over 2}\nu_c} P_\nu d\nu$ is well determined by our observations, we find a synchrotron radiation energy density:
$$u_{syn} = {P_{syn} \over 4\pi R_{43}^2 c} = 7.44\,10^{-4} ~ \delta^{\alpha-2} (R/R_{43})^{-2} ~{\rm J\,m^{-3}} $$
For $\delta \simeq 1$ and $R=R_{43}$ we get $u_{syn} \la u_B$. 
On the other hand, Centaurus A has been detected in $\gamma -$rays \citep{ThompsonEGRET_1995,Steinle_etal1998}, showing a broad luminosity peak at $\nu_{IC} = 3 10^{19}$\,Hz with $\nu F_\nu = 5\,10^{-13} {\rm  Wm^{-2}} = 5\,10^{13}$\,Hz\,Jy, that is $\sim 3$ times more powerful then the synchrotron peak in Fig.\,\ref{f:spectrum}. As we found electron energies $\langle \gamma \rangle$ of a few hundred (depending on $R/R_{43}$), which could up-scatter synchrotron photons from $\nu_{syn} \simeq 3\,10^{13}$\,Hz to $\nu_{IC} = 2\gamma^2 \nu_{syn} \ga 10^{19}$\,Hz, it seems plausible to follow the standard interpretation of this second peak as {\it synchrotron self Compton} radiation \citep[SSC,][]{JonesOdellStein_1974,ChCaCe_2001}.  The observed SSC luminosity $P_{IC}$ requires
\begin{equation}
\label{equ:ratio_u}
{u_{syn} \over u_B} = 0.744 ~ \delta^\alpha\: \left( {R \over R_{43}} \right)^{-10} = {P_{IC} \over P_{syn} } \ga 3.
\end{equation}
Obviously, condition (\ref{equ:ratio_u}) is fulfilled if the radius of the synchrotron core, $R_c$, is slightly smaller than the observed value at 43\,GHz: $R_c = 0.87\, R_{43} = 1.26\: 10^{14}$m, that is well within the 20\% error estimated for $R_{43}$. As observationally $u_{syn}$ is much better determined than $u_B$, we will use the parameters of the synchrotron core derived from (\ref{equ:ratio_u}) in the following. They 
are summarized in Table~\ref{t:parameters}. The here derived parameters of the synchrotron source are in good qualitative agreement with those derived by \citet{ChCaCe_2001} on the basis of their SSC model assuming standard variability arguments and thus demonstrate that the basic properties of the synchrotron core do not rely too much on our detailed assumptions. However, it should be noted that the extension of the $\gamma-$ray peak into the X--ray region ($5\,10^{17} \dots 2\,10^{18}$\,Hz) is significantly steeper, $\nu^{-0.7}$ \citep{Evans_etal2004},  than that expected from a {\it pure} SSC model. Thus an additional source of X--rays might be present ({\it cf.} section \ref{ss:ionisation}).

\subsubsection{Relation to the radio jet}
\label{ss:syn_derived}

To investigate the nature of the synchrotron source more closely, it is worthwhile to pursue the consequences of our essential measurements -- namely the exact values of the cutoff frequency $\nu_c$ and the power-law slope $\alpha$ -- even further: the most natural explanation for the high frequency cutoff is, that at the corresponding particle energy $\gamma_c$ the radiation loss time $\tau_{loss}(\gamma_c)$ exactly equals the acceleration time scale $\tau_{acc}(\gamma_c)$:
\begin{eqnarray*}
\tau_{acc}(\gamma_c) \equiv {\gamma_c \over d\gamma / dt} = 
\tau_{loss} &= 3.8\,10^6 {\rm s} \left( u_B + u_{syn} \over {1\, {\rm J\,m^{-3}} } \right)^{-1} \gamma_c^{-1}\\ &= 3.44\, 10^5 {\rm s} = 4.0\, {\rm  days,}
\end{eqnarray*}
where we used $u_B,~u_{syn}$, and $\gamma_c$ from Table~\ref{t:parameters}. The first thing to notice is, that $\tau_{acc} = 4$\,days agrees well with the variability time scale $\tau_{var} \simeq 1$\,day observed in the 100\,MeV range \citep{Kinzer_etal95}.
Second, $c\tau_{acc}(\gamma_c) = 1.03\,10^{14}$\,m, is of the same order as
the source radius $R_c$. Assuming an energy-independent $\tau_{acc}$ particles have to travel at least a distance $l_{acc} \simeq \log_2(\gamma_c/200) c\tau_{acc}(\gamma_c) = 5.5\, 10^{14}$\,m to be accelerated from $\gamma = 200$ to $\gamma = \gamma_c$. As $l_{acc} \ga 4 R_c$, the particles have to cross the source several times or gain a considerable amount of their energy on the way to the source.

\begin{figure}
\centering
  \includegraphics[width=90mm]{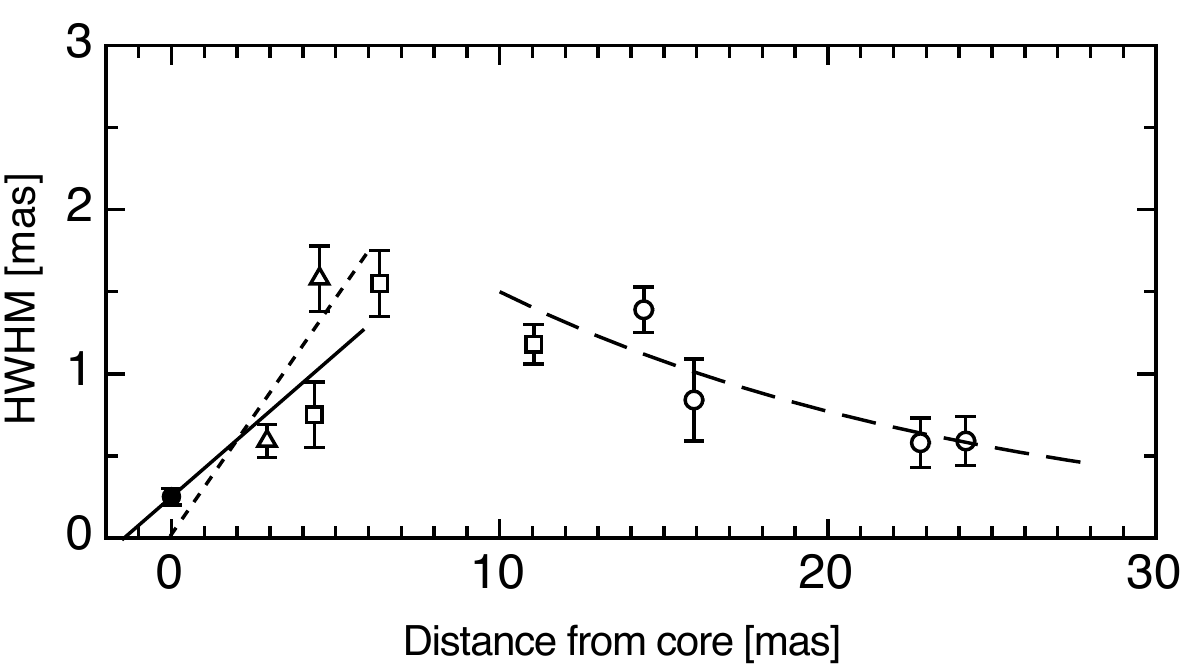}
      \caption{Half width of the VLBA jet as a function of the distance from the radio core ($\bullet$). The values have been derived from clearly resolved components C1 ($\circ$), C2 ($\Box$), and C3 ($\triangle$) on the 8.4 GHz maps from \citet{Tingay_etal98}. Two alternatives for the jet opening angle are shown: either the radio "core" represents the innermost (stationary) knot of the radio jet some $\sim 1.4$\,mas (0.026\,pc) from the origin (solid line, half opening angle 10\degr ) or it is located at the core proper (dashed line, maximum opening angle 16\degr ). Our measurements have been confirmed recently by \citet{Horiuchi_etal2006}.}
         \label{f:jetwidth}
\end{figure}

This leads to the question of the nature of the synchrotron source and its distance from the central black hole. The most likely interpretation is, that the source represents the ''base'' of the radio jet, that is the innermost
point at which electrons reach highly relativistic energies.
From the width of the VLBA jet (Fig.\,\ref{f:jetwidth}) it seems that the
jet is expanding freely out to a distance of about 5 mas ($= 0.1$\,pc $= 3\,10^{15}$\,m) from the VLBA core. In this region the jet half opening angle is between 10\degr\ and 16\degr . This leads to an upper limit of $d \le 0.026$\,pc$ \simeq 6\,R_c$ for the distance of the synchrotron source from the core proper.

The slope of the synchrotron powerlaw $\alpha = -0.36$ corresponds to an electron spectrum $n(\gamma) \propto \gamma^{-q}$ with $q = 1.72$. This is considerably flatter than the standard value expected from Fermi acceleration at a strong, non-relativistic shock ($q = 2$). However, it has been
demonstrated by various authors \citep{KirkSchneider_1987, KirkHeavens_1989} that first order Fermi acceleration at (oblique) relativistic shocks could produce power-law slopes between $q=1.6$ and $q = 2$. An alternative way to produce such flat electron spectra could be provided by relativistic current sheets \citep{Kirk_Texas2005}. In this context, it is instructive to check whether the magnetic field and relativistic particle energy density are close to equipartition ($u_B \simeq u_{e\pm}$) as it seems to be the case in the
terminal shocks (hot spots) of extended radio jets \citep{Meise_etal89}. With
the parameters in Table~\,\ref{t:parameters} we find for the energy in electrons and positrons:
\begin{eqnarray*}
u_{e\pm} &=& 1000\,m_ec^2  n_{1000} \int_{0.1}^{8.5}{g^{1-q}dg} 
= 2.4\;10^{-4}\; \delta^{-2}R_c^{4\alpha-7}\; {\rm J\,m^{-3}}\\ &\simeq &0.74\times u_B, 
\end{eqnarray*}
where $g \equiv \gamma/1000$ and we assume $\gamma_{min} = 100$. So, unless a lot more energy is stored in relativistic protons, the synchrotron core does not deviate far from equipartition.

To summarize, we interpret the synchrotron "core" of Centaurus A as the
innermost point of its relativistic outflow, at which interaction with the surrounding medium leads to the onset of efficient particle acceleration  within the jet flow. At our present knowledge, it cannot be decided whether this "visible base" of the jet is marked by an internal shock or magnetic reconnection phenomena in the relativistic flow. In any case, it seems likely, that the onset of radiation from the jet is connected to deceleration of the flow to $\Gamma_{jet} \la 2.5$.
As relativistic particles of moderate energies ($\gamma < 1000$) suffer smaller synchrotron losses, one might speculate that they are advected downstream with the jet flow, thus providing the "seed particles" which are required for efficient shock acceleration in the parsec-scale radio jet.

It is worth to note, that recent observations of the kiloparsec jet with the Spitzer Space Telescope \citep{Quillen_etal2006,Brookes_etal2006} have established that its synchrotron spectrum shows a radio-to-infrared power-law
with $\alpha = -0.72$ which extends at least to $\nu = 10^{14}$\,Hz. Assuming an equipartition magnetic field of $\simeq 3$\,nT one derives that  electrons in the kiloparsec jet have to be accelerated to energies $\gamma_{max} \ga 10^6$ ({\it i.e.} $100 \times \gamma_c$ of the synchrotron core). This might indicate
that the maximum energy scales with the size of the acceleration region.

\subsection{Circum-nuclear dust emission from the parsec disk}
\label{ss:thermaldust}

\begin{figure}
\centering
 \includegraphics[width=92mm]{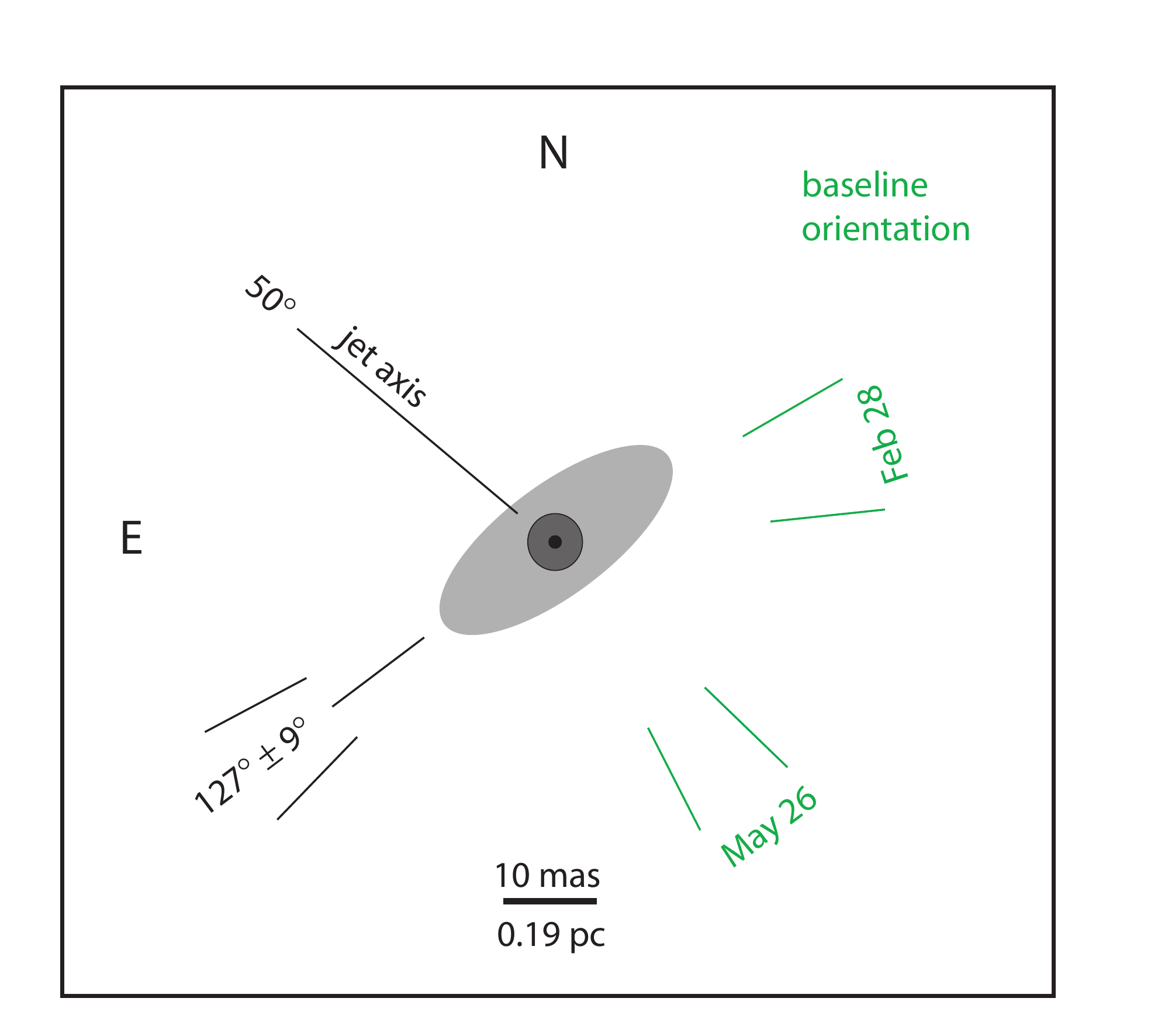}
      \caption{Sketch of our model for the mid-infrared emission from the inner parsec of Centaurus A. We identify the unresolved point source of $<6$\,mas FWHM (dark grey) with the VLBI core (FWHM $= 0.5\pm 0.1$\,mas, indicated as black dot). It is surrounded by an elongated structure of dust emission (light grey) the major axis of which is orientated along $P.A. = 127\degr \pm 9\degr$ as inferred from the orthogonal baselines observed on May 26. From the visibilities observed with two baselines on February 28 we derive a major axis length of
      about 30 mas. Note that the major axis orientation is consistent with being
      perpendicular to the radio jet axis, and the axis ratio can be explained by a thin disk the axis of which is inclined by $\sim 66\degr$ with respect to our line-of-sight. }
         \label{f:sketch}
\end{figure}

As pointed out in section \ref{s:results} our current -- very limited -- coverage of the $uv$-plane leads us to the conclusion, that the center of Centaurus A is essentially unresolved along $P.A. \simeq 40\degr$ but shows a clear indication for an extended component along $P.A. \simeq 120\degr$. Until future interferometric observations with other baselines allow us to constrain better the size and shape of the extended component, we simply assume that the visibility $V^{Feb28}(\lambda) = 0.8 - 0.04 (\lambda - 8\,\mu{\rm m})$ along
$P.A. = 108\degr \pm 12\degr$ is caused by the superposition of the unresolved synchrotron core and a well resolved, inclined disk, the major axis of which must be orientated roughly perpendicular to $P.A. = 37\degr\pm 9\degr$, along which we find $V(\lambda) \simeq 1$. The size of the disk is poorly confined by the present observations but needs to be
$\ga 30$\,mas ($= 0.57$\,pc) at $\lambda = 13\,\mu$m, in order to be consistent
with our simple two-component model. As there might be a marginal decrease in the visibility along $P.A. \simeq 40\degr$ towards the longest wavelengths, only an upper limit of $\sim 12$\,mas can be given for the projected width of the disk. Figure \ref{f:sketch} sketches this interpretation. Note that within the
current uncertainties the major axis of the disk could well be orientated exactly perpendicular to the direction of the parsec scale radio jet at $P.A.(jet) = 50\degr$ \citep{Tingay_etal98} and could represent an inclined thin disk, the axis of which is aligned with the radio axis at $50\degr < \theta < 70\degr$ with respect to our line-of-sight ({\it cf.} Table~\ref{t:parameters}).

From the visibility $V^{Feb28}(\lambda)$ and the extinction corrected flux values in Table~\ref{t:spectrum} we derive $F_{\rm disk}(8.3\mu{\rm m}) = (0.21\pm 0.10)$\,Jy and $F_{\rm disk}(12.6\mu{\rm m}) = (0.71\pm 0.20)$\,Jy, respectively.\footnote{When assuming an unresolved core and a well resolved disk, $F_{\rm disk}$ is related to the core flux $F_{0,\nu}$ by $ F_{\rm disk}(\lambda) =
( {1 \over V(\lambda)} -1)\, F_{\nu,0}(\lambda)$.} Interpreting this steep rise towards long wavelengths as the Wien tail of a blackbody spectrum from warm dust leads to a dust temperature of $T \simeq 240$ K. With this temperature we derive a rough estimate of the bolometric power emitted by the dust: $P_{\rm dust} \ga 3\; 10^{34}$\,W. 
Since the dust disk seems too thin to cover more than $1\pi$ steradian (seen from the central accretion disk), we conclude, that a heating power $P_{heat} \ge 10^{35}$\,W is required to explain the apparent dust emission. As we will discuss in section\,\ref{ss:ionisation} the optical-UV power radiated by a nuclear accretion disk is insufficient to heat the dust. Thus other radiation sources must illuminate the dust disk. It seems that the most likely source for its heating is provided by X-ray radiation. 
Regarding the size of the dust disk, it is instructive to calculate the dust sublimation (inner) radius for the required heating power $P_{heat} \ga 10^{35}$\,W and a sublimation temperature of 1500\,K:

$$r_{in} = 1.3\,{\rm pc} \left({P_{heat} \over 10^{39}\,{\rm W} }\right)^{1/2}
         \ga 0.013\,{\rm pc}. $$

As this corresponds to $< 1$\,mas, that is $< 1/20$ of our resolution, it is hard to determine how much the innermost parts of the dust disk could contaminate
the flux from the unresolved core. In any case,
we conclude, that the amount of dust emission from the central parsec of Centaurus A is most likely limited by the available illuminating radiation.
Thus it is impossible to determine the gas and dust content of the innermost parsec from mid-infrared observations. 

\subsection{Thermal radiation from the core of Centaurus A}
\label{ss:ionisation}

Even when allowing for the Doppler effect along the jet axis, the maximum photon energy   $\delta_0 h\nu_c < 1.5\, {\rm  eV} \ll 13.6$\,eV, of the synchrotron radiation from the core of Centaurus A cannot ionize hydrogen. Thus, the observation of narrow emission lines from the nucleus of Centaurus~A requires a thermal ionization source.
A crude upper limit for the ionizing flux from a thermal source (accretion disk?) within the core of Centaurus A might be obtained by assuming that the entire excess at $\lambda = 0.814\,\mu$m, $F_{\nu,0} - F_{syn} \simeq 9$\,mJy, is due to a thermal source\footnote{Note: the extremely faint "core" observed at $\lambda = 0.55\,\mu$m by \citet{Marconi_etal00}, when corrected for $A_V = 14$\,mag, would have an intrinsic flux of $F_{\nu,0} \simeq 36$\,mJy, much higher than any reasonable accretion disk
spectrum could account for. We conclude therefore, that this flux does not come from the core proper.} and that this source has an intrinsic spectrum resembling that of typical accretion disks in type 1 AGN, $F_\nu \sim \nu^{-0.6}$ out to $\nu = 6\:10^{16}$\,Hz. This yields $P_{16} \simeq 3\,10^{34}$\,W, that is only $0.1\%$ of that of a typical Seyfert nucleus or $4\,10^{-5} \times L_{Edd}$,  the Eddington luminosity of the black hole. At such low luminosity (and correspondingly low accretion rate), the accretion flow onto the black hole will not occur via a thin accretion disk, but in the form of an Advection Dominated Accretion Flow \citep[ADAF,][]{NarayanYi_1995, Narayan_etal1998}.
A further possibility to estimate the amount of thermal radiation can be
based on the observed Br$_\gamma$ flux of $6\:10^{-19}$\,Wm$^{-2}$ within a $3\times 3\, \sq\arcsec$ aperture (Neumayer et al. {\it in prep.}).
Assuming, that the gas disk intercepts $0.7\pi$ steradian\footnote{That is an angular range of $\pm 10\degr$ or a full disk height of 7\,pc at half the radius of the disk, $r/2 = 20$\,pc.}
of an isotropically emitted ionizing radiation and that every recombining H atom emits $0.004\: {\rm Br}\gamma$ photons on average \citep[case A,][]{Osterbrock_1989} one finds that $P_{Lyc} = 1.3\,10^{34}$\,W has to be emitted isotropically in the Lyman continuum. 
Thus, the above derived estimate for the thermal power, $P_{16}$ seems  sufficient to provide the ionizing flux which is needed to explain the observed line emission around the core of Centaurus A. 

As outlined in section \ref{ss:thermaldust}, at least $10^{35}$\,W of (isotropic) luminosity are required to account for the emission of the dust disk. Obviously this cannot be provided by the UV radiation of such a low luminosity accretion flow. Significantly more power is available in X-rays. So far, we have discussed the high frequency emission from the core of Centaurus~A only in terms of the SSC model.
%
Indeed, the low frequency tail of the inverse Compton (IC) radiation between $3\,10^{15}$ and $10^{17}$\,Hz, 
expected from the SSC models might provide some ionizing photons, which due to their hard spectrum could lead to rather high ionization states. As obvious from the $\delta_0$ values in Table~\ref{t:parameters}, we expect Doppler boosting of this IC radiation between a factor of 5 and 36 along the jet axis! In fact, recent observations with the adaptive optics spectrograph SINFONI
reveal that coronal lines like [SiVI] are aligned along the jet axis (van der Werf et al. \& Neumayer et al., {\it in prep.}).
In addition,
Chandra and XMM observations \citep{Evans_etal2004} between 2 and 7 keV ($\nu = 5$ to $17\times 10^{17}$\,Hz)  
detected an absorbed ($N_H \simeq 10^{23}\,{\rm cm}^{-2}$) continuum source with a $\nu^{-0.7}$ spectrum and a flux of $F_\nu = 63\,\mu$Jy at 1 keV ($2.42\;10^{17}$\,Hz) which the authors interpret as emission from an accretion disk. As this spectrum is considerably steeper than that expected for pure SSC emission of the synchrotron core ($\nu^{-0.36}$), one needs to  consider that part of the X-ray flux is of thermal origin: we obtain a rough estimate of the thermal contribution by integrating the $\nu^{-0.7}$ power-law between $10^{17}$ and $10^{19}$\,Hz and subtracting the SSC contribution. This yields a thermal X-ray luminosity of $P_{X,th} \ga 1\,10^{35}$\,W.  
To summarize, we conclude that the X-ray luminosity seems to be {\it just} able to heat the dust. Including the X-ray flux the thermal luminosity of the nucleus in Centaurus~A is $P_{th} \simeq 1.5\,10^{-4} \times L_{Edd}$. 

It should be mentioned that the high HI column would argue for $A_V \simeq 50$\,mag if the standard interstellar conversion $A_V/N_H = 5\times10^{-22}$\,mag cm$^2$ would be assumed. Although this seems in conflict with our value $A_V \simeq 14$\,mag, one should note that for $A_V = 14$\,mag the ratio $A_V/N_H = 1.4\times10^{-22}$\,mag cm$^2$ lies well in the range that has been found in other FR I radio galaxies \citep{Balmaverde_etal2006}. 

\subsection{Comparison with other radio sources}
\label{ss:others}

We now compare our findings on Centaurus A with other AGN which host a similarly massive black hole. It is evident, that even the closest and least luminous
Seyfert 2 galaxies (Circinus, NGC 1068) contain nuclear dust concentrations ("tori") which radiate $10\dots100\times$ more powerful in the mid-IR than Centaurus A. As discussed in section \ref{ss:thermaldust} this  mainly might be explained by the lack of a heating source, while the total amount of cold dust in the inner parsecs can hardly be constrained. So certainly, Centaurus A is not a "well hidden" Seyfert galaxy.  

Instead, it shares many properties of nearby FR I radio galaxies: Morphology and luminosity of its parsec to kiloparsec jets is well in the range observed for
nearby FR Is. As in other FR I galaxies \citep{Balmaverde_etal2006,Evans_etal2006}, its nuclear X-ray emission is produced
at least partly at the base of the radio jet. Most of the dust extinction towards the core occurs in dust structures on scales of 50 to thousands of parsec. However, the existence of a narrow-line region which exhibits high ionization lines and our new evidence for a very compact nuclear dust disk (0.6 pc diameter) are features of Centaurus A, which are untypical for FR I radio galaxies. It is worth noting, that the small dust luminosity of Centaurus A places it among the "mid-IR weak" radio galaxies which comprise about half of a sample of FR II galaxies observed by \citet{Ogle_etal2006} with the Spitzer Space Observatory.

The most unique feature of Centaurus A is its rather powerful synchrotron core, the spectrum of which peaks around $10^{14}$ Hz. Even the much more powerful
radio galaxy M\,87 which hosts a 50 times more massive black hole cannot compete with Centaurus A in this respect. Since relativistic beaming cannot account for the difference\footnote{Most authors believe that the M\,87 jet is within about
20\degr\ of our line of sight. This would argue for a significantly higher Doppler factor ($\delta \gg 1$) in M\,87 than inferred for Centaurus A.}, we would like to argue that the luminous synchrotron core is an intrinsic property of Centaurus A. It is attractive to speculate, that the more-than-average amount of dust and gas in the innermost parsec, as established by our detection of extended mid-infrared emission, could play an important role in building up the strong internal shock at $d < 0.026$ pc which is capable of converting a significant fraction of the out-flowing kinetic energy into relativistic particles already such close to the core.  

Finally, we would like to discuss the term "mis-directed" BL Lac which has been
used by several authors to explain the unique properties of the non-thermal core in Centaurus A. In the framework of the unified scheme \citep{UrryPadovani_1995}  any FR I radio galaxy qualifies as  "mis-directed BL Lac". The question is, however, whether the {\it special properties} of the synchrotron core in Centaurus A are typical for BL Lac objects. The strongest argument against this is the bulk Lorentz factor $\Gamma_{jet} < 2.5$, which seems outside the range $5 < \Gamma_{jet} < 32$ derived by \citet{UrryPadovani_1995} for radio-selected BL Lac objects. Accordingly, neither the
fact that the synchrotron spectrum reaches to frequencies as high as $10^{14}$\,Hz, nor the $\gamma-$ray emission, nor the variability time scale seem to be caused by relativistic beaming (we derive a Doppler factor $\delta \simeq 1$).
Rather they seem to be a consequence of violent interaction between the relativistic outflow and the surrounding medium, which occurs much closer to the central black hole (at $r < 10^4\,R_S$) than in typical FR I radio galaxies.
However, we cannot exclude, that the parsec jet of Centaurus A contains a relativistic ($\Gamma > 5$) "spine", which leaves no imprint on any of the observations obtained so far.

\section{Conclusions}

Our interferometric observations of Centaurus A in the N-band ($8 < \lambda < 13\,\mu$m) provide strong evidence that its mid-infrared emission is dominated by an unresolved synchrotron core. The size of this core is most likely given by the size of $R_{43} \simeq 0.01$\,pc, derived from VLBI observations at 43\,GHz. Additionally, the observations with interferometric baselines orientated roughly perpendicular to the parsec scale radio jet revealed an extended component which naturally can be interpreted as a geometrically thin, dusty disk, the axis of which coincides with the radio jet. Its diameter is about 0.6\,pc. It contributes between 20\% (at short wavelengths, $\lambda \simeq 8\mu$m) and 40\% (at $\lambda \simeq 13\mu$m) to the nuclear flux from Centaurus A and contains dust which is heated up to about 240 K.

We demonstrate, that assuming an extinction $A_V = (14\pm 2)$\,mag, all flux measurements of the core between radio and near infrared frequencies can be fitted nicely by a synchrotron spectrum, although there is evidence for variability. The spectrum is characterized by a rather flat power-law $F_\nu \sim \nu^{-0.36}$ which cuts off above $\nu_c = 8\; 10^{13}$\,Hz and becomes optically thick below $\nu_1 \simeq 4.5\; 10^{10}$\,Hz. Following the usual interpretation of the $\gamma-$ray emission from Centaurus A as Synchrotron Self Compton (SSC) radiation we derive a magnetic field strength of 26 $\mu$T and an maximum energy of relativistic electrons of $\gamma_c = E_c/m_ec^2 = 8500$. With these parameters we derive an acceleration time scale of $\tau_{acc} = 4$~days, which is in  good agreement with the fastest flux variations, observed at X-ray frequencies. We point out, however, that the spectral slope at X-ray frequencies does not fit into the SSC model, but requires an additional X-ray emission process.

Our SSC model argues for a Doppler factor $\delta \simeq 1$ which -- together with the jet-counter jet ratio of the radio jets on parsec scale -- results in an upper limit $\Gamma_{jet} < 2.5$. Such a low bulk Lorentz factor does not fit to the concept of a "mis-directed BL Lac object", unless
there exists a highly relativistic "spine", which has no observable signature.

Finally, we try to estimate the thermal luminosity from the accretion flow  around the black hole in Centaurus A:  Taking the observed excess at $\lambda < 1\,\mu$m above the synchrotron spectrum as signature of a thermal core component and assuming that part of the X-ray flux is of thermal origin, we find a thermal power $P_{th} \simeq 1.3\;10^{35}\,{\rm W} \simeq 1.5\,10^{-4} \times L_{Edd}$. Although this is at least two orders of magnitude below the value in highly radiation-efficient accreting AGN ({\it e.g.} Seyfert galaxies), it is still substantially higher than the values which are typical for FR I radio galaxies. Nevertheless, it remains in the range predicted for Advection Dominated Accretion Flows \citep[ADAF,][]{Narayan_etal1998} where most of the accretion energy is lost unseen or is channeled into kinetic jet power. 
At the present state of knowledge, one cannot decide whether its relatively high thermal luminosity indicates that
Centaurus A currently is undergoing a transition between low and high radiative efficiency \citep[\textit{cf.} ][]{Falke_unified2004} or whether this intermediate state can be maintained over long periods of time. In any case, the special properties of Centaurus A -- the closest and best resolved active galactic nucleus -- seem to provide a counter-example against simple unified schemes, which try to explain FR I radio galaxies and BL Lacertae objects by orientation effects alone.

\begin{acknowledgements}
      We thank Paul van der Werf for access to the SINFONI data before publication. Several aspects of the paper could be clarified in discussions with John Kirk and Max Camenzind. We are particularly grateful to the referee, Yuan Feng, whose constructive criticism helped us clarify and improve this paper substantially.
\end{acknowledgements}

\bibliography{MIDI_general,CenA_reference,AGN_Meise,AGN_general}
\bibliographystyle{aa}

\end{document}